\documentclass[pdflatex,sn-mathphys-num]{sn-jnl}% Math and Physical Sciences Numbered Reference Style
\usepackage{graphicx}%
\usepackage{multirow}%
\usepackage{amsmath,amssymb,amsfonts}%
\usepackage{amsthm}%
\usepackage{mathrsfs}%
\usepackage[title]{appendix}%
\usepackage{xcolor}%
\usepackage{textcomp}%
\usepackage{manyfoot}%
\usepackage{booktabs}%
\usepackage{algorithm}%
\usepackage{algorithmicx}%
\usepackage{algpseudocode}%
\usepackage{listings}%

\usepackage{threeparttable}
\usepackage{multibib}

\newcites{met}{References (continued)}
\newcites{supp}{References (continued)}

\bibliographystylemet{sn-mathphys-num}
\bibliographystylesupp{sn-mathphys-num}

\DeclareRobustCommand{\mainmetcite}[2]{%
  \citetext{\citealp{#1},\citealpmet{#2}}%
}

\DeclareRobustCommand{\mainsuppcite}[2]{%
  \citetext{\citealp{#1},\citealpsupp{#2}}%
}

\newcommand{\ion}[2]{#1\,{\sc #2}}

\theoremstyle{thmstyleone}%
\theoremstyle{thmstyletwo}%

\theoremstyle{thmstylethree}%

\begin{document}

% \title[Article Title]{Article Title}

% %%=============================================================%%
% %% GivenName	-> \fnm{Joergen W.}
% %% Particle	-> \spfx{van der} -> surname prefix
% %% FamilyName	-> \sur{Ploeg}
% %% Suffix	-> \sfx{IV}
% %% \author*[1,2]{\fnm{Joergen W.} \spfx{van der} \sur{Ploeg} 
% %%  \sfx{IV}}\email{iauthor@gmail.com}
% %%=============================================================%%

% \author*[1,2]{\fnm{First} \sur{Author}}\email{iauthor@gmail.com}



% \affil*[1]{\orgdiv{Department}, \orgname{Organization}, \orgaddress{\street{Street}, \city{City}, \postcode{100190}, \state{State}, \country{Country}}}

% \affil[2]{\orgdiv{Department}, \orgname{Organization}, \orgaddress{\street{Street}, \city{City}, \postcode{10587}, \state{State}, \country{Country}}}

% \affil[3]{\orgdiv{Department}, \orgname{Organization}, \orgaddress{\street{Street}, \city{City}, \postcode{610101}, \state{State}, \country{Country}}}
% \title[Multiphase outflows in a gas-rich merger 500 Myr after the Big Bang]{Multiphase outflows in a gas-rich merger 500 Myr after the Big Bang from JWST ultra-deep spectroscopy}
\title[An emerging baryon cycle in a galaxy 500 million years after the Big Bang]{An emerging baryon cycle in a galaxy 500 million years after the Big Bang}
% lead authors
\author[1,2]{\fnm{Shengzhe} \sur{Wang}}

\author*[1,2,3]{\fnm{Xin} \sur{Wang}}
\email{xwang@ucas.ac.cn}

% coauthors providing code or specific analysis
\author[1]{\fnm{Hang} \sur{Zhou}}

\author[4]{\fnm{Zhijie} \sur{Qu}}

\author[5]{\fnm{Zhaozhou} \sur{Li}}

\author[1]{\fnm{Yuxuan} \sur{Pang}}

\author[1]{\fnm{Qianqiao} \sur{Zhou}}

\author[6,7]{\fnm{Shouyi} \sur{Wang}}

% coauthors providing comments
\author[5]{\fnm{Yangyao} \sur{Chen}}

\author[8]{\fnm{Yuguang} \sur{Chen}}

\author[9]{\fnm{Karl} \sur{Glazebrook}}

\author[9]{\fnm{Glenn G.} \sur{Kacprzak}}

\author[10]{\fnm{Nicha} \sur{Leethochawalit}}

\author[11]{\fnm{Houjun} \sur{Mo}}

\author[12]{\fnm{Themiya} \sur{Nanayakkara}}

\author[13]{\fnm{Huiyuan} \sur{Wang}}

% alphabetic orders
\author[14]{\fnm{Weida} \sur{Hu}}

\author[1]{\fnm{Xunda} \sur{Sun}}

\author[2]{\fnm{Chao-Wei} \sur{Tsai}}

\author[2]{\fnm{Hu} \sur{Zhan}}

\affil*[1]{School of Astronomy and Space Science, University of Chinese Academy of Sciences (UCAS), Beijing 100049, China}

\affil[2]{National Astronomical Observatories, Chinese Academy of Sciences, Beijing 100101, China}

\affil[3]{Institute for Frontiers in Astronomy and Astrophysics, Beijing Normal University, Beijing 102206, China}

\affil[4]{Department of Astronomy, Tsinghua University, Beijing 100084, China}

\affil[5]{School of Astronomy and Space Science, Nanjing University, Nanjing 210093, China}

\affil[6]{Department of Astronomy, School of Physics, Peking University, Beijing 100871, People's Republic of China}

\affil[7]{Kavli Institute for Astronomy and Astrophysics, Peking University, Beijing 100871, China}

\affil[8]{Department of Physics, The Chinese University of Hong Kong, Shatin, N.T., Hong Kong, China}

\affil[9]{Centre for Astrophysics and Supercomputing, Swinburne University of Technology, Hawthorn, VIC 3122, Australia}

\affil[10]{National Astronomical Research Institute of Thailand (NARIT), Mae Rim, Chiang Mai 50180, Thailand}

\affil[11]{Department of Astronomy, University of Massachusetts Amherst, Amherst, MA 01003, USA}

\affil[12]{Sydney Institute for Astronomy, School of Physics, The University of Sydney, Sydney, NSW 2006, Australia}

\affil[13]{Department of Astronomy, University of Science and Technology of China, Hefei 230026, China}

\affil[14]{Shanghai Astronomical Observatory, Chinese Academy of Sciences, 80 Nandan Road, Shanghai 200030, China}


\abstract{The emergence of stellar feedback as a regulator of galaxy growth marks a fundamental transition in cosmic history \citep{Tumlinson_2017}.
At early times, rapid gas accretion and collapse may induce intense star formation before feedback becomes effective, producing feedback-free starbursts \citep{Dekel_2023,Li_FFB_2024}.
When and how such bursts subsequently develop into self-regulated baryon cycles remain observationally unknown. Here we show that Gz9p3, a merging galaxy at $z=9.311$, is caught in this transition only 500 million years after the Big Bang \citep{Boyett_2024}. Deep JWST spectroscopy reveals a substantial neutral-gas reservoir along its merger-driven tidal structure and a multiphase outflow.
Fine-structure absorption provides the first direct measurement of the electron density of the cool outflowing gas at high redshift ($\approx\,17\,{\rm cm^{-3}}$), yielding a mass-loading factor among the highest yet measured for galaxies of comparable stellar mass.
The emergence of such efficient feedback after an intense burst is consistent with the delayed onset of feedback expected in feedback-free starburst models. The cool outflowing gas is unlikely to escape the host halo, implying that much of this metal-enriched material may remain available for future recycling through the circumgalactic medium. Gz9p3 therefore provides an early view of a baryon cycle being established through the interplay of merger-driven gas redistribution, bursty star formation and stellar feedback, suggesting that feedback-regulated recycling was already shaping galaxy growth during the epoch of reionization.
}

\maketitle

% \section{Introduction}\label{sec1}

Galaxy growth is governed by the cycling of baryons between gas accretion, star formation, stellar feedback and the surrounding circumgalactic medium (CGM) \citep{Tumlinson_2017}.
In the first few hundred million years, however, the high gas densities and rapid accretion rates of young galaxies may allow gas to collapse and form stars on timescales shorter than those required for feedback to disrupt the star-forming gas reservoir. 
JWST has revealed a large number of unexpectedly luminous galaxies at these epochs, motivating the feedback-free starburst (FFB) models in which intense starbursts convert gas into stars with high efficiency before stellar feedback becomes dynamically effective \citep{Naidu_2022,Harikane_2023,Donnan_2023,Finkelstein_2024,Sun_2023,Dekel_2023,Li_FFB_2024}.
Such a feedback-free phase cannot persist indefinitely. Photoionization and stellar winds begin to act within a few Myr, followed by core-collapse supernovae, ultimately removing gas from star-forming regions and establishing feedback regulation on timescales of order $10$--$30\,\mathrm{Myr}$ \citep{Leitherer_1999_starburst,Hopkins_2012,Chevance_2020,Eldridge_2022}.
A key untested prediction of this picture is therefore a brief transition from FFB to a feedback-regulated baryon cycle.
When this transition first occurred, how efficiently feedback redistributed gas, and whether the expelled material escaped or remained available for recycling are unknown.

Galactic outflows provide a direct observational tracer of the onset of feedback regulation. At lower redshifts, rest-frame ultraviolet (UV) absorption spectroscopy has revealed multiphase winds through blueshifted transitions spanning a wide range of ionization states, constraining their velocities, ionic columns and covering fractions ($C_{\rm f}$) \citep{Heckman_1990,Veilleux_2005,Steidel_2010,Rupke_2018,Rudie_2019}. JWST has extended measurements of ionized outflows to the Epoch of Reionization through broad H$\alpha$ and [\ion{O}{III}] emission \citep{Cooper_2025,Xu_2025,Carniani_2024}. Yet whether these winds were already capable of regulating galaxy growth remains uncertain.
Their baryonic impact is commonly quantified by the mass-loading factor ($\eta=\dot{M}_{\rm out}/{\rm SFR}$), i.e., the ratio between mass outflow rate ($\dot{M}_{\rm out}$) and star-formation rate (SFR).
However, estimates of $\dot{M}_{\rm out}$ depend sensitively on gas density, geometry, covering fraction and spatial extent---quantities that are rarely measured directly at high redshift \citep{Cooper_2025,Carniani_2024,Nakane_2026}. In particular, the poorly constrained density of the outflowing gas remains a major uncertainty in determining how efficiently early galaxies exchanged baryons with their surroundings.

A decisive test of the emergence of feedback regulation requires connecting the gas reservoir that fuels star formation to %the material expelled by feedback and 
the physical conditions of the outflows. Such simultaneous constraints are rarely available at cosmic dawn. Gz9p3 provides this opportunity at $z=9.311$, when the Universe was only $\sim500\,\mathrm{Myr}$ old. %after the Big Bang.
It is a luminous, gas-rich merger with $\log(M_\star/M_\odot)=9.2^{+0.1}_{-0.2}$, two central luminosity peaks and an extended stellar tail \citep{Boyett_2024}. Previous analyses have revealed intense recent star formation and a rapidly evolving star-formation history (SFH), including a short recent burst \citep{Boyett_2024,Chen_2026}. 
% Its disturbed morphology and exceptional JWST spectroscopic coverage allow the neutral-gas reservoir, the multiphase outflow seen in absorption, and the ionized outflow traced in emission to be studied within the same system.

% Here we use ultra-deep JWST rest-frame UV spectroscopy to show that Gz9p3 is undergoing a transition from burst-dominated growth to strong feedback regulation. Spatially resolved damped Ly$\alpha$ (DLA) absorption reveals a substantial neutral-gas reservoir associated with its extended tidal structure, while absorption-line diagnostics and broad [\ion{O}{III}] emission reveal a multiphase galactic outflow. Fine-structure absorption directly constrains the electron density ($n_{\rm e}$) of the cool outflowing gas, substantially reducing a major uncertainty in its mass-loading factor, which is among the highest measured for galaxies of comparable stellar mass. The emergence of such efficient feedback following an intense recent burst is consistent with the delayed onset of feedback expected in FFB models. The cool outflow is unlikely to escape the host halo, suggesting that feedback primarily redistributes gas into the CGM environment rather than permanently ejecting it. Gz9p3 therefore captures an early stage in which rapid stellar growth, feedback-driven gas redistribution and recycling are becoming coupled into a self-regulated baryon cycle.

Here we combine spatially resolved damped Ly$\alpha$ absorption (DLA) measurements of the neutral-gas reservoir across Gz9p3 and its extended tidal structure with direct constraints on the density and mass loading of a multiphase outflow. The emergence of efficient feedback shortly after the recent burst is consistent with the delayed onset expected in FFB models, while the cool outflowing gas appears unlikely to escape the host halo. Gz9p3 therefore captures an early stage in the establishment of a self-regulated baryon cycle, in which rapid stellar growth is followed by feedback-driven redistribution of gas that remains available for subsequent recycling.

% Throughout this work, we adopt a flat $\Lambda$CDM cosmology with $H_0 = 70~\mathrm{km~s^{-1}~Mpc^{-1}}$, $\Omega_{\mathrm{M}} = 0.3$, and $\Omega_{\Lambda} = 0.7$.

% \section{Results}\label{sec2}

% \subsection{Multiphase Galactic Outflow}

% \subsection{}

\section*{Efficient feedback after a recent starburst}

\begin{figure}
\centering
\includegraphics[width=\textwidth]{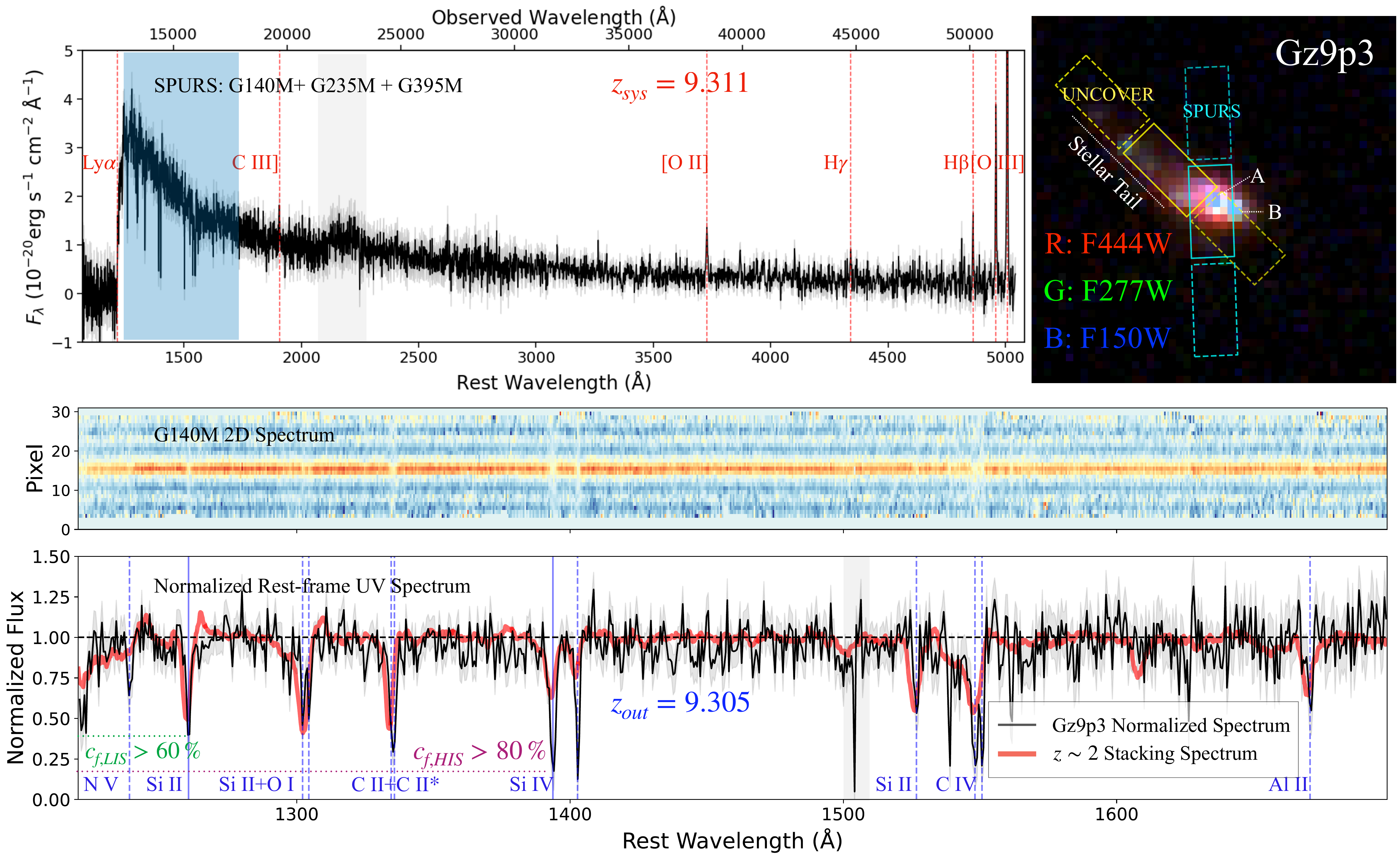}
\vspace{-1em}
\caption{\textbf{Spectroscopic signatures of multiphase galactic outflows in Gz9p3.}
\textbf{Top left:} 
The ultra-deep medium-resolution JWST/NIRSpec spectrum covers wavelengths from the rest-frame UV to the optical of Gz9p3, with the prominent emission features marked. The blue shaded region is expanded in the lower two panels, while the gray shaded region marks a detector defect.
\textbf{Top right:} Three-color JWST/NIRCam image revealing the disturbed morphology of Gz9p3. The two central luminosity peaks are labeled as components A and B, while the extended low-surface-brightness structure is marked as the stellar tail. The yellow and cyan boxes indicate the NIRSpec/MSA footprints of the UNCOVER and SPURS observations, respectively.
\textbf{Middle:} Two-dimensional NIRSpec spectrum covering the rest-frame UV.
\textbf{Bottom:} Continuum-normalized rest-frame UV spectrum of Gz9p3 (black) compared with the stacked spectrum of $z\sim2$ star-forming galaxies \citep{Du_2018} (red). 
% Among the comparison samples available across different redshifts, this stack exhibits an absorption-line velocity offset most similar to that observed in Gz9p3 and is therefore shown to facilitate a direct comparison of their outflow signatures.
Low-ionization state (LIS) and high-ionization state (HIS) absorption lines are labeled. The absorption-line centroids indicate an outflow redshift of $z_{\rm out}=9.305$, corresponding to $v_{\rm out}\approx160\,{\rm km\,s^{-1}}$. Saturated absorption lines imply lower limits on the covering fractions of $C_{\rm f,LIS}>60\%$ and $C_{\rm f,HIS}>82\%$.
}
\label{fig:Gz9p3_spec}
\end{figure}

The ultra-deep JWST spectroscopy of Gz9p3 reveals a multitude of 
% a multiphase galactic outflow only $\sim500\,\mathrm{Myr}$ after the Big Bang (Fig.~\ref{fig:Gz9p3_spec}). 
low-ionization state (LIS) and high-ionization state (HIS) absorption lines, coherently blueshifted by $\sim150$--$200\,\mathrm{km\,s^{-1}}$ (Fig.~\ref{fig:Gz9p3_spec}).
The saturated \ion{Si}{II} and \ion{Si}{IV} absorption imply covering-fraction limits of $C_{\rm f,LIS}>0.6$ and $C_{\rm f,HIS}>0.8$, respectively. 
Despite their similar kinematics, the substantially different covering fractions rule out a single homogeneous absorber and instead point to a multiphase outflow, with cooler low-ionization gas embedded within a more extended highly ionized medium. Voigt-profile modeling constrains the absorber kinematics and ionic columns and places the absorbing gas within $R<5\,\mathrm{kpc}$ of the central star-forming region (Methods). Independently, broad [\ion{O}{III}]~$\lambda5007$ emission shows little velocity offset from systemic, consistent with contributions from both the near and far sides of the outflow (Methods).

The detection of strong \ion{N}{V} absorption raises the possibility of an active galactic nucleus (AGN) contribution to the HIS phase. An AGN-dominated origin is, however, disfavored by the absence of broad hydrogen recombination lines, the X-ray non-detection, and the moderate, coherent velocities of the HIS absorption \citep{Brazzini_2025,Kraemer_2001}. A direct origin in active massive-star winds is also not favored, as stellar \ion{N}{V} and \ion{C}{IV} wind features are typically broad P-Cygni profiles reaching velocities of order $10^3\,\mathrm{km\,s^{-1}}$ \citep{Rickard_2022}, unlike the relatively narrow, high $C_{\rm f,HIS}$ absorption observed in Gz9p3. Instead, the HIS column-density ratios overlap the regime occupied by collisionally ionized and non-equilibrium cooling gas (Extended Data Fig.~\ref{fig:HIS_1}), while their column densities and moderate velocity offsets are more consistent with galactic outflows and halo gas than with the high-velocity absorbers characteristic of luminous AGN (Extended Data Fig.~\ref{fig:HIS_2}). We therefore interpret the HIS absorption as tracing feedback-heated interfaces or cooling gas associated with the multiphase outflow, although a moderate, obscured, or X-ray-weak AGN cannot be excluded.

\begin{figure}
\centering
\includegraphics[width=\columnwidth]{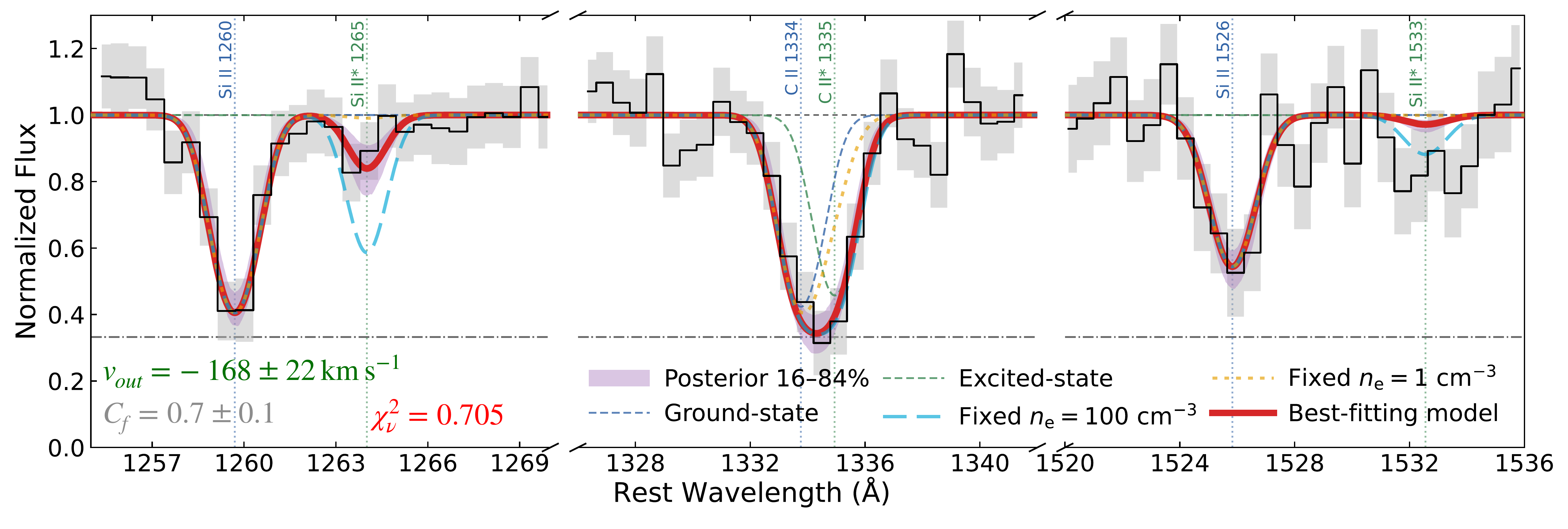}
\vspace{-1em}
\caption{\textbf{Direct measurement of
the electron density ($n_{\rm e}$) of the outflowing gas from the \ion{Si}{II} and \ion{C}{II} fine-structure absorption lines.}
The continuum-normalized spectra (black) and their $1\sigma$ uncertainties (gray) are shown for the \ion{Si}{II} and \ion{C}{II} resonance transitions and their corresponding fine-structure absorption lines. The blue and green dotted lines indicate the expected centroids of the ground-state and excited-state transitions, respectively. The red curve shows the best-fitting joint partial-covering model, while the purple shaded region denotes the 16th--84th percentile posterior interval from the MCMC analysis. 
% All transitions are fitted simultaneously using a single physical model with common $v_{\rm out}$, $C_{\rm f}$, and $n_{\rm e}$. 
The relative strengths of the resonance and fine-structure absorption lines directly constrain the excited-level populations and hence the electron density of the outflowing gas. The best-fitting model --- with a reduced chi-square of $\chi_\nu^2=0.705$ --- self-consistently reproduces both the saturated resonance lines and the much weaker fine-structure transitions, providing the first direct measurement of $\log(n_{\rm e}/{\rm cm}^{-3})=1.22^{+0.44}_{-0.71}$ for high-redshift galactic outflows.
}
\label{fig:CII_ne}
\end{figure}

Quantifying the baryonic impact of the outflow requires the gas density, which remains one of the dominant uncertainties in high-redshift outflow measurements. In Gz9p3, the relative populations of the ground and fine-structure levels of \ion{C}{II} and \ion{Si}{II} provide a direct density diagnostic. We jointly model the \ion{C}{II}, \ion{C}{II}$^*$, \ion{Si}{II}, and \ion{Si}{II}$^*$ absorption systems, linking their level populations through collisional excitation (Fig.~\ref{fig:CII_ne}; Methods). The resulting fit yields $\log(n_{\rm e}/{\rm cm}^{-3})=1.22^{+0.44}_{-0.71}$.

The direct density constraint substantially reduces one of the principal uncertainties in converting the absorption-line measurements into a baryonic mass flux. We infer a mass-loading factor of $\eta=12.3^{+21.7}_{-9.9}$ by combining the measured $n_{\rm e}$ with the outflow kinematics, $C_{\rm f}$, projected continuum-emitting area, and the H$\beta$-based SFR (Methods). Unlike conventional outflow estimates that rely on an assumed outflow radius or dynamical timescale, our absorption-line calculation uses the directly constrained gas density and therefore does not require an assumed outflow radius, opening angle, or dynamical time. 
This estimate assumes that the hydrogen in the cool phase gas is fully ionized, thereby minimizing the total hydrogen density inferred from $n_{\rm e}$. The resulting $\eta$ should therefore be regarded as a lower limit for the ionized outflow alone and does not include any contribution from neutral gas.
Broad [\ion{O}{III}] emission provides a complementary view of the outflow, as the spatially integrated line profile can include contributions from both the approaching and receding material. However, applying the conventional emission-line prescription with a commonly adopted outflow density of $n_{\rm e}=380\,\mathrm{cm^{-3}}$ yields $\eta^{[\ion{O}{III}]}=0.13^{+0.08}_{-0.06}$ \citep{Cooper_2025,Carniani_2024}, nearly two orders of magnitude below the absorption-line estimate. This discrepancy illustrates the strong dependence of conventional emission-line outflow measurements on the assumed gas density, and shows that adopting empirical density prescriptions can substantially underestimate the strength of feedback when the physical conditions of the outflow are not directly constrained.

%%%%%% Shengzhe Wang 8.7

\begin{figure}
\centering
\includegraphics[width=\textwidth]{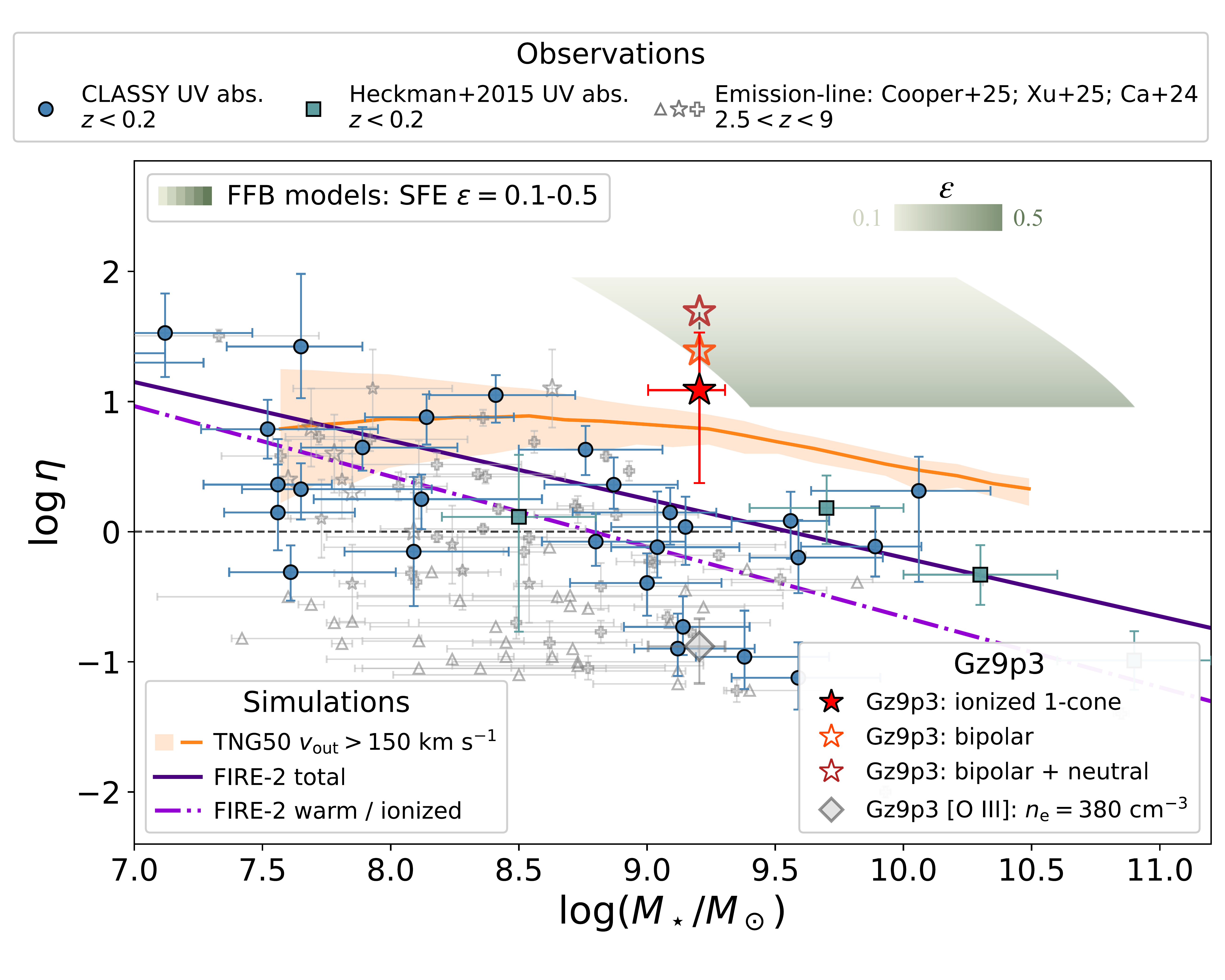}
\vspace{-1em}
\caption{\textbf{Outflow mass-loading factor ($\eta$) as a function of galaxy stellar mass.}
The UV absorption lines trace only the foreground, approaching side of the cool ionized outflow projected against the UV-bright stellar continuum, for which we measure a mass-loading factor of $\eta=12.3^{+21.7}_{-9.9}$ (red filled star).
We multiply this value by a factor of two to account for a bipolar outflow geometry (i.e. the orange open star).
Under a conservative multiphase assumption in which the neutral and ionized phases contribute equal masses to the outflow \citep{Fluetsch_2019,Fluetsch_2021,Davies_2024}, the resulting estimate (brown open star) lies $\sim1$--$2$ orders of magnitude above simulation predictions \citep{Nelson_2019,Pandya_2021} and most previous observational measurements.
Here observations include UV absorption measurements on local star-forming galaxies and CLASSY analogs \citep{Heckman_2015, Chisholm_2017, Xu_2022}, as well as emission-line measurements from JWST/NIRSpec at high redshift \citep{Cooper_2025, Xu_2025, Carniani_2024}. However we caution that the latter method results in a severe underestimate of $\eta$ for Gz9p3 (i.e., the gray diamond).
Interestingly, the FFB models with a star-formation efficiency ($\varepsilon$) ranging from 0.1 to 0.5 can reproduce the apparent outflow mass loading measured in Gz9p3, supporting a scenario in which the delayed onset of strong feedback produces massive multiphase galactic outflows; the construction of the dark-green shaded region is described in Methods.
}
\label{fig:eta}
\end{figure}

Gz9p3 is extreme on the $\eta$--$M_\star$ plane (Fig.~\ref{fig:eta}). The directly measured one-sided ionized loading factor already lies above the UV absorption-line measurements at comparable stellar mass. Accounting for the two-sided outflow geometry indicated by the broad [\ion{O}{III}] emission further increases the inferred loading relative to emission-line measurements. For comparison with simulations, we conservatively include an equal contribution from neutral gas, yielding the ``bipolar+neutral'' estimate; this remains above the commonly adopted TNG and FIRE predictions at $M_\star\simeq10^{9.2}\,\mathrm{M_\odot}$. Because the density of the cool ionized phase is measured directly and no correction for projection of the outflow velocity is applied, these estimates remain conservative. The unusually large mass loading of Gz9p3 therefore requires a physical explanation beyond conventional uncertainties in outflow modeling.

The rapidly evolving SFH of Gz9p3 provides a natural explanation for its extreme observed mass loading.
The reconstructed SFH indicates that the galaxy experienced a short, intense burst approximately $10$--$20\,\mathrm{Myr}$ before the epoch of observation, during which its SFR was about an order of magnitude higher than that traced by H$\beta$ over the most recent $\sim10\,\mathrm{Myr}$ \citep{Chen_2026}. The absorbing outflow is not observed instantaneously after the burst: gas expelled by strong feedback requires time to propagate away from the star-forming regions and to cover a significant fraction of the UV continuum.
The conservative upper limit on the outflow radius ($R<5\,\mathrm{kpc}$), together with the observed projected velocity ($v_{\rm out}\simeq160\,\mathrm{km\,s^{-1}}$), places an upper limit of $\sim30\,\mathrm{Myr}$ on the propagation timescale, leaving the recent starburst as a viable epoch for launching the observed outflow.
The wind observed today can therefore retain the imprint of the preceding burst while its mass-loading factor is normalized by the substantially lower current SFR, naturally producing an enhanced apparent $\eta$.

This temporal sequence provides a natural connection to the FFB models. In these models, rapid gas collapse permits a highly efficient episode of star formation --- with an efficiency $\varepsilon$ as high as 50\% --- before stellar feedback becomes dynamically effective, followed by massive galactic outflows driven by strong feedback from the FFB stellar populations \citep{Dekel_2023,Li_FFB_2024,Chen_FFB_2026}.
We extend the wind framework of \citet{Li_FFB_2024} by accounting for the observed SFH of Gz9p3 and the temporal offset between the burst and the outflow.  
Adopting the observed decline in star formation, corresponding to a fiducial ratio $f_{\rm burst}\simeq0.1$ between the current and burst star-formation rates, the geometry- and phase-corrected loading factors correspond to characteristic burst efficiencies of $\varepsilon\simeq0.17$--$0.28$ under our fiducial assumptions (Methods). 
The precise inferred $\varepsilon$ remains dependent on the adopted hot--cool phase decomposition, outflow geometry and completeness of the observed gas phases, and should therefore be interpreted as a physically motivated constraint within the FFB framework rather than a unique determination of the burst efficiency. Nevertheless, the combination of an intense recent burst followed by an unusually high present-day mass-loading factor is consistent with the delayed emergence of stellar feedback expected as the system transitions from FFB into a feedback-regulated phase.

Gz9p3 therefore appears to be caught as strong stellar feedback emerges after an intense recent burst. The key question is whether this feedback permanently removes baryons from the system or instead redistributes them within the halo, leaving them available for subsequent recycling.

\section*{Gas redistribution rather than escape}

Efficient feedback does not necessarily imply that baryons are permanently removed from a galaxy. The observed LIS outflow velocity remains below the estimated escape velocity ($v_{\rm esc}$) of the host halo. We evaluate $v_{\rm esc}$ at $2\,\mathrm{kpc}$, approximately the projected radius of the system \citep{Cooper_2025}, and at $5\,\mathrm{kpc}$, the conservative upper limit on the radial distance of the LIS absorbing gas inferred from the UV absorption analysis (Methods). For the plausible halo-mass range of Gz9p3, we obtain $v_{\rm esc}\simeq350$--$650\,\mathrm{km\,s^{-1}}$ and $\simeq290$--$560\,\mathrm{km\,s^{-1}}$, respectively, both exceeding the observed line-of-sight velocity of the cool outflow. Although projection, halo-mass uncertainty and subsequent acceleration prevent a definitive dynamical classification, the measurements indicate that a substantial fraction of the cool outflowing material remains gravitationally bound.

\begin{figure}[htbp]
\centering
\includegraphics[width=\textwidth]{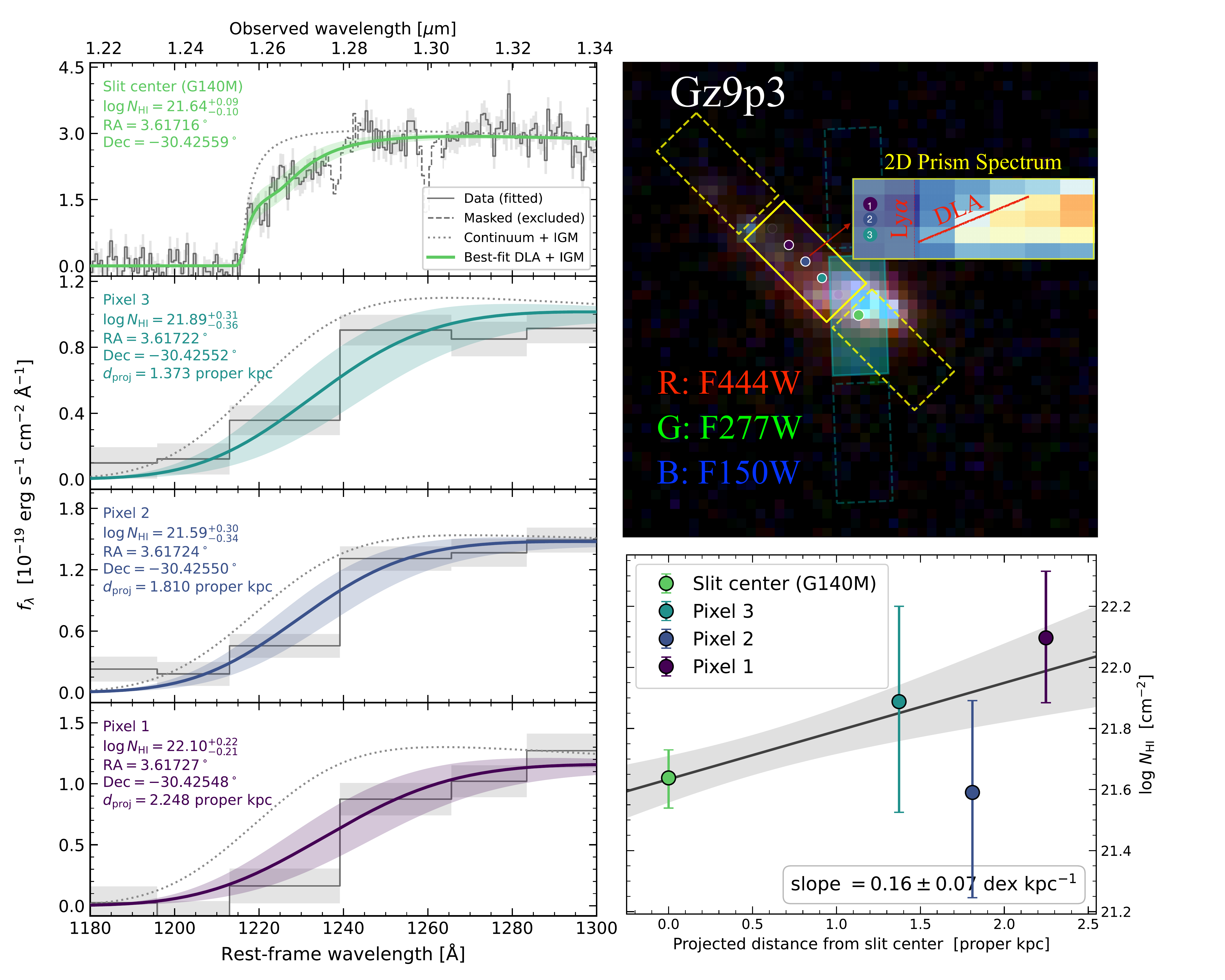}
\vspace{-1em}
\caption{\textbf{Spatially resolved DLA fitting reveals neutral-gas redistribution in Gz9p3.}
\textbf{Left:}
% Ly$\alpha$ damping-wing fits at the G140M slit center and three spatial pixels along the tidal structure.
Ly$\alpha$ damping-wing fits at the G140M slit center and at three spatial pixels extracted from the prism spectrum along the tidal structure (Pixels 1--3).
The shaded regions show the $1\sigma$ uncertainty ranges from the MCMC fitting. The posterior distributions from the joint fit to the four sightlines are shown in Supplementary Fig.~\ref{fig:DLA_corner_1}.
\textbf{Top right:} Three-color NIRCam image of Gz9p3.
As in Fig.~\ref{fig:Gz9p3_spec}, the yellow and cyan boxes indicate the NIRSpec/MSA footprints of the prism and G140M spectroscopy, respectively.
% The yellow slit indicates the sky coverage of the MSA shutters used for the prism spectroscopy. 
We also show the original two-dimensional prism spectrum in the inset, in which the DLA absorption exhibits clear spatial variation.
% The cyan apertures mark the sky coverage of the MSA shutters used for the G140M observations highlighted in Figure~\ref{fig:Gz9p3_spec}.
\textbf{Bottom right:} Combining the DLA joint-fitting results from the four sightlines, we derive the variation of $\log N_{\rm HI}$ as a function of projected distance from the main galaxy center and perform a simple linear fit. The $N_{\rm HI}$ shows a weak increasing trend along the tidal tail, possibly caused by dense \ion{H}{I} clumps within the tidal structure.
}
\label{fig:DLA}
\end{figure}

Spatially resolved DLA provides a complementary view of the neutral-gas reservoir in Gz9p3 (Fig.~\ref{fig:DLA}). We jointly model four sightlines spanning the galaxy center and the extended stellar-tail direction over a projected scale of $\simeq2.5\,\mathrm{kpc}$. Although the absolute normalization of $N_{\rm HI}$ remains model-dependent because of degeneracies between local absorption and the intergalactic medium (IGM) damping wing, the relative spatial variation is more robust within the joint framework (Methods). The fitted sightlines span $\log(N_{\rm HI}/{\rm cm}^{-2})\simeq21.6$--$22.1$, with the outer sightlines along the stellar tail showing comparable or higher $N_{\rm HI}$ than the galaxy center. Gz9p3 therefore contains a substantial neutral-gas reservoir associated with its extended stellar structure.

The neutral-gas distribution closely follows the disturbed morphology of Gz9p3. The system contains two central luminosity peaks and an extended stellar tail (Fig.~\ref{fig:Gz9p3_spec}), along which pixel-by-pixel spectral energy distribution (SED) modeling indicates in-situ star formation \citep{Boyett_2024}. The spatial coincidence between the stellar tail, enhanced $N_{\rm HI}$, and ongoing star formation suggests that the merger has redistributed neutral gas over kiloparsec scales, potentially providing the fuel for star formation outside the main stellar body. Similar configurations are observed in nearby gas-rich mergers, where substantial atomic-gas reservoirs can reside in tidal structures \citep{van_der_Hulst_1979,Yun_1994,Hibbard_1996,Hibbard_2001,Duc_2013}.

The observed $N_{\rm HI}$ gradient does not uniquely require an absolute enhancement of neutral gas in the tail: the lower central column density may also reflect gas consumption or clearing associated with the recent starburst and feedback. A projected feedback-driven shell cannot be excluded with the limited sightline coverage, but the spatial correspondence between high $N_{\rm HI}$, the stellar tail and in-situ star formation naturally favors merger-driven tidal redistribution as the dominant origin of the large-scale neutral-gas structure.

Taken together, the apparently bound cool outflow and the extended neutral-gas reservoir point to redistribution rather than wholesale baryon loss from Gz9p3. The merger redistributes neutral gas across the system, while stellar feedback displaces metal-enriched material from the central star-forming regions into the CGM without necessarily ejecting it from the halo. Such material can therefore remain available for subsequent recycling, linking merger-driven gas redistribution and stellar feedback into an emerging baryon cycle.

\bibliography{Gz9p3}% common bib file

% ============================================================================
% Gz9p3: replacement Methods + Extended Data + Supplementary Information
% Paste this block after \bibliography{Gz9p3} in the current manuscript.
% Remove the old block beginning with \newpage before \section*{Methods} and
% ending with \bibliographysupp{Gz9p3}. Keep the existing \end{document}.
% The preamble and main text are intentionally not included here.
% ============================================================================

\clearpage
\setcounter{figure}{0}
\renewcommand{\figurename}{Extended Data Fig.}
\renewcommand{\thefigure}{\arabic{figure}}
\renewcommand{\theHfigure}{ExtendedDataFigure.\arabic{figure}}
\setcounter{table}{0}
\renewcommand{\tablename}{Extended Data Table}
\renewcommand{\thetable}{\arabic{table}}
\renewcommand{\theHtable}{ExtendedDataTable.\arabic{table}}

% \DeclareRobustCommand{\mixedcite}[2]{%
%   \citetext{\citealp{#1},\citealpsupp{#2}}%
% }

\section*{Methods}

%1
\subsection*{Observations and data reduction}
\label{app:data_reduction}

Gz9p3 is located at RA $=3.6171694$, Dec $=-30.4255494$ and has a gravitational lensing magnification of $\mu\simeq1.66$ \citep{Boyett_2024}. Our analysis is primarily based on JWST NIRSpec MSA spectroscopy from the medium-resolution SPURS observations (JWST GO-9214 \citepmet{Tang_2026}) and the prism observations from UNCOVER (JWST GO-2561 \citepmet{Bezanson_2024_uncover}). The SPURS data were obtained with G140M/F100LP ($29.2\,\mathrm{hr}$), G235M/F170LP ($7.9\,\mathrm{hr}$), and G395M/F290LP ($2.9\,\mathrm{hr}$), providing rest-frame coverage from $1000\,$\text{\AA} to $5050\,$\text{\AA} at $R\sim1000$. The UNCOVER PRISM/CLEAR observation ($2.6\,\mathrm{hr}$) covers a similar wavelength range at lower spectral resolution and provides the spatial information used for pixel-by-pixel extraction.

We retrieved the uncalibrated rate products (\texttt{*\_rate.fits}) from the Mikulski Archive for Space Telescopes and processed them using \texttt{MSAEXP} version 0.9.2 \citepmet{msaexp}, the JWST Science Calibration Pipeline version 1.16.0, and CRDS context \texttt{jwst\_1312.pmap}. The processing includes WCS initialization, slit identification, flat-fielding, pathloss correction, photometric calibration, and drizzling of the two-dimensional spectra onto a rectified common grid. For the prism observations, we modeled and subtracted the background using source-free regions within the two-dimensional shutters to preserve the spatial structure of the extended source. For the medium-resolution observations, the background was removed using standard nodding subtraction. Ancillary imaging, X-ray analysis, and other available JWST observations are described in Supplementary Section~``\hyperref[supp:ancillary]{Ancillary data and additional JWST observations}''.

%2
\subsection*{Spatially resolved DLA fitting}
\label{app:DLA}

We model the Ly$\alpha$ absorption profiles of Gz9p3 using the NIRSpec MSA prism and G140M/F100LP spectra with a joint DLA framework following \citet{Chen_2026}. In this high-redshift regime, the model includes both local \ion{H}{I} absorption associated with Gz9p3 and damping-wing attenuation from the neutral IGM along the line of sight. The prism slit intersects the extended stellar-tail structure, from which we extract spectra at three spatial positions along the slit. These spectra are combined with the G140M/F100LP spectrum of the galaxy center, yielding four spatially resolved sightlines across a projected extent of $2.5\,\mathrm{kpc}$ (Fig.~\ref{fig:DLA}).

For each sightline, the intrinsic rest-frame UV continuum is described by a local power law, $F_\lambda \propto \lambda^\beta$, and the local neutral hydrogen column density, $N_{\rm HI}$, is allowed to vary independently. Because all four sightlines probe the same galaxy and nearly the same foreground IGM, we assume common global IGM parameters, including the volume-averaged neutral fraction, $\bar{x}_{\rm HI}$, and the distance from the galaxy to the onset of the neutral IGM.

This joint spatially resolved approach is designed to isolate relative sightline-to-sightline variations in the local \ion{H}{I} column from the global IGM attenuation. The absolute normalization of $N_{\rm HI}$ remains model-dependent, owing to degeneracies among continuum placement, local DLA absorption, the IGM neutral fraction, and the distance to the neutral IGM. Across the four sightlines, the fitted column densities span $\log(N_{\rm HI}/{\rm cm}^{-2})\simeq21.6$--$22.1$.

However, our main conclusion does not rely on the absolute column density scale alone. Within this common modeling framework, the outer prism sightlines along the stellar tail show higher $N_{\rm HI}$ than the central G140M/F100LP sightline, indicating an enhanced neutral-gas reservoir associated with the extended stellar-tail structure. The details of the continuum modeling, priors, and MCMC implementation are described in Supplementary Information section~``\hyperref[supp:DLA]{Additional details of the DLA fit}''.

\subsection*{Absorption-line measurements and Voigt-profile fitting}
\label{app:voigt}

The ultra-deep rest-frame UV spectroscopy from SPURS covers a rich set of absorption-line features in Gz9p3, including low-ionization state (LIS) transitions such as \ion{Si}{II}, \ion{O}{I}, \ion{C}{II}, \ion{C}{II}$^*$, and \ion{Al}{II}, as well as high-ionization state (HIS) transitions including \ion{N}{V}, \ion{Si}{IV}, and \ion{C}{IV}. After normalizing the continuum, we measure the rest-frame equivalent widths (EW) of these absorption lines and perform saturation-line diagnostics to constrain $C_{\rm f}$ of saturated transitions (see Supplementary Information ~``\hyperref[supp:EW]{Equivalent-width analysis}''). Using a curve-of-growth argument, we find that \ion{Si}{II}\,$\lambda1260$ and \ion{Si}{IV}\,$\lambda1393$ are saturated. The residual absorption depths of these saturated transitions then provide lower limits on the covering fractions of the LIS and HIS gas, yielding $C_{\rm f,LIS}>0.6$ and $C_{\rm f,HIS}>0.8$, respectively.

We perform MCMC-based Voigt-profile fitting to the absorption lines to place conservative constraints on the ionic column densities and velocity offsets of different ions (see  Supplementary Information~``\hyperref[supp:voigt]{Additional details of the Voigt-profile fitting}''). For the column-density measurements, we adopt $C_{\rm f}=1$ and allow two velocity components to account for unresolved substructures. We report lower limits on the ionic column densities rather than treating them as precise measurements, and use them mainly to compare the relative abundance of different ionic species (see Extended Data Table~\ref{tab:ion}).

\begin{table*}[htbp]
\centering
\footnotesize
\begin{threeparttable}
\caption{Best-fit ionic column densities and centroid velocities.}
\label{tab:ion}
\tabcolsep 12pt
\begin{tabular}{lcc}
\toprule
Ion & $\log N$ (dex) & $v$ (km\,s$^{-1}$) \\
\midrule
Si\,II & $15.1^{+1.0}_{-0.3}$ & $-158 \pm 17$ \\
O\,I   & $15.5^{+1.4}_{-0.3}$ & $-158 \pm 17$ \\
C\,II\tnote{a} & $15.9^{+1.1}_{-0.5}$ & $-168 \pm 20$ \\
Al\,II & $13.5^{+1.6}_{-0.4}$ & $-177 \pm 56$ \\
N\,V   & $14.5^{+0.3}_{-0.3}$ & $-185 \pm 68$ \\
Si\,IV & $14.9^{+1.4}_{-0.2}$ & $-174 \pm 15$ \\
C\,IV  & $15.2^{+0.7}_{-0.2}$ & $-205 \pm 26$ \\
\bottomrule
\end{tabular}
\begin{tablenotes}
\footnotesize
\item Column densities are in logarithmic units of cm$^{-2}$. Except for C\,II, they are conservative lower-limit estimates from multi-component fitting. Uncertainties are 68\% highest-posterior-density (HPD) intervals estimated with kernel-density estimate (KDE). Centroid velocities are from the single-component fits and are relative to systemic.
\item[a] The C\,II column density is from a single-component fit with free covering fraction and includes C\,II$^*$.
\end{tablenotes}
\end{threeparttable}
\end{table*}

\subsection*{Fine-structure absorption and electron density}
\label{app:CII_ne}

As shown in Fig.~\ref{fig:CII_ne}, the absorption feature near \ion{C}{II}\,$\lambda1334$ is noticeably broader and deeper than the saturated \ion{Si}{II}\,$\lambda1260$ line. At the spectral resolution of the G140M spectrum, $R\sim1000$, the ground-state transition \ion{C}{II}\,$\lambda1334$ cannot be cleanly separated from the fine-structure transition \ion{C}{II}$^*$\,$\lambda1335$. The excess absorption around this wavelength can therefore be interpreted in two ways: either the \ion{C}{II} absorption contains an additional velocity component not seen in the other LIS transitions, or a significant fraction of the absorption arises from \ion{C}{II}$^*$\,$\lambda1335$.

We consider the second interpretation more likely. The other LIS transitions, including \ion{Si}{II}, \ion{O}{I}, and \ion{Al}{II}, do not show clear evidence for an additional velocity component at the same location. If the excess absorption were instead produced by a separate \ion{C}{II} component, it would have to be strong in carbon while being weak or absent in Si, O, and Al, requiring an unusual abundance or ionization pattern. Moreover, this interpretation would require an additional \ion{C}{II} component redshifted by approximately $100\,\mathrm{km\,s^{-1}}$ relative to the systemic velocity, corresponding to a possible inflowing component. No analogous redshifted absorption is seen in the other LIS transitions. Such a carbon-only inflowing component is therefore not independently supported by the data. By contrast, the \ion{C}{II}$^*$ interpretation naturally accounts for the excess absorption at the expected fine-structure wavelength within the same outflow velocity structure. We therefore interpret the excess absorption near \ion{C}{II}\,$\lambda1334$ primarily as evidence for significant \ion{C}{II}$^*$ absorption.

The combination of strong \ion{C}{II}$^*$ absorption and the non-detection of significant \ion{Si}{II}$^*$ absorption provides a useful density diagnostic for the cool LIS gas. We model the \ion{C}{II}, \ion{C}{II}$^*$, \ion{Si}{II}, and \ion{Si}{II}$^*$ absorption systems with a single velocity component. Specifically, we jointly fit \ion{Si}{II}\,$\lambda\lambda1260,1526$, \ion{Si}{II}$^*$\,$\lambda\lambda1265,1533$, \ion{C}{II}\,$\lambda1334$, and \ion{C}{II}$^*$\,$\lambda1335$. We exclude \ion{Si}{II}\,$\lambda1304$ from this fit because it is strongly degenerate with \ion{O}{I}\,$\lambda1302$.

To connect the fine-structure level populations to the gas density, we use \texttt{ChiantiPy}, the Python interface to the \texttt{CHIANTI} atomic database \citepmet{Dere_1997,Del_Zanna_2021,Dere_2023}, to compute the expected column-density ratios $N(\ion{C}{II})/N(\ion{C}{II}^{*})$ and $N(\ion{Si}{II})/N(\ion{Si}{II}^{*})$ as a function of $n_{\rm e}$, assuming $T_{\rm e}=10^4\,\mathrm{K}$ \citepmet{Mao_2025}. We then perform a six-parameter MCMC fit with $n_{\rm e}$, $\log N(\ion{C}{II})$, $\log N(\ion{Si}{II})$, the Doppler parameter $b$, the velocity offset $\Delta v$, and the covering fraction $C_{\rm f}$ all treated as free parameters. Allowing $C_{\rm f}$ to vary is important for this analysis, because the $n_{\rm e}$ constraint is driven by the relative absorption depths of the resonance and fine-structure lines.

\begin{figure}[htbp]
\centering
\includegraphics[width=0.8\textwidth]{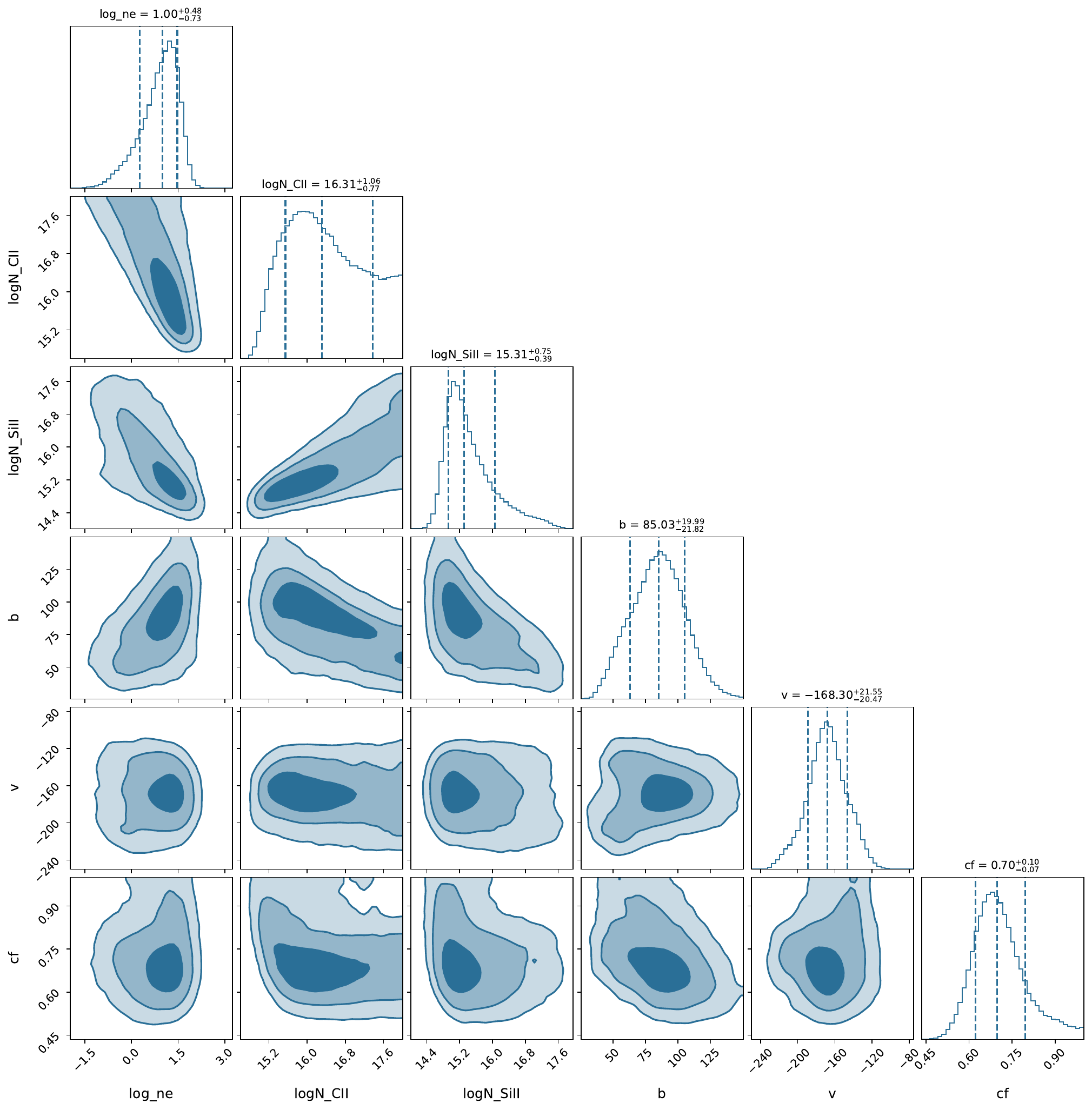}
\caption{Posterior distributions from the joint MCMC fit to the C\,II, C\,II$^*$, Si\,II, and Si\,II$^*$ absorption systems.}
\label{fig:ne_MCMC}
\end{figure}

The posterior distribution is shown in Extended Data Fig.~\ref{fig:ne_MCMC}. This fit yields $\log(n_{\rm e}/{\rm cm}^{-3})=1.22^{+0.44}_{-0.71}$. We report the posterior mode as the fiducial value and the 68\% HPD interval as the uncertainty. This differs from the percentile values shown on the corner plot, which indicate the 16th, 50th, and 84th percentiles of the marginalized posterior.

The inferred electron density also depends weakly on the assumed electron temperature, particularly in the low-density limit where the collisional excitation rates become more sensitive to $T_{\rm e}$. To quantify this effect, we repeated the analysis assuming $T_{\rm e}=15{,}000\,\mathrm{K}$ and $20{,}000\,\mathrm{K}$. These assumptions increase the inferred density by only $\simeq0.09$ and $0.20$ dex, respectively, relative to the fiducial value at $T_{\rm e}=10{,}000\,\mathrm{K}$. These shifts are smaller than the statistical uncertainty of the measurement, $\log(n_{\rm e}/{\rm cm}^{-3})=1.22^{+0.44}_{-0.71}$. Moreover, the inferred density lies around $n_{\rm e}\sim10\,\mathrm{cm^{-3}}$, where the temperature dependence is not expected to dominate the fine-structure excitation balance. We therefore adopt $T_{\rm e}=10{,}000\,\mathrm{K}$ as our fiducial assumption.

Resonant scattering and fluorescent re-emission could partially fill the fine-structure troughs \citepmet{Gazagnes_2023}. However, fluorescent emission in local absorption-selected outflows is generally weak, narrow and kinematically distinct from the broad blueshifted absorption \citepmet{Xu_2023}, and reprocessed photons may emerge outside the spectroscopic aperture \citepmet{Wang_2020_fluo}. We detect no significant \ion{Si}{II}$^*$ or \ion{C}{II}$^*$ emission in either the one- or two-dimensional spectra. Tests using nearby CLASSY galaxies are presented in Supplementary Section~``\hyperref[supp:density]{Electron-density robustness tests and local comparison}''.

\subsection*{[\ion{O}{III}] emission-line decomposition}
\label{app:OIII}

\begin{figure}[htbp]
\centering
\includegraphics[width=\columnwidth]{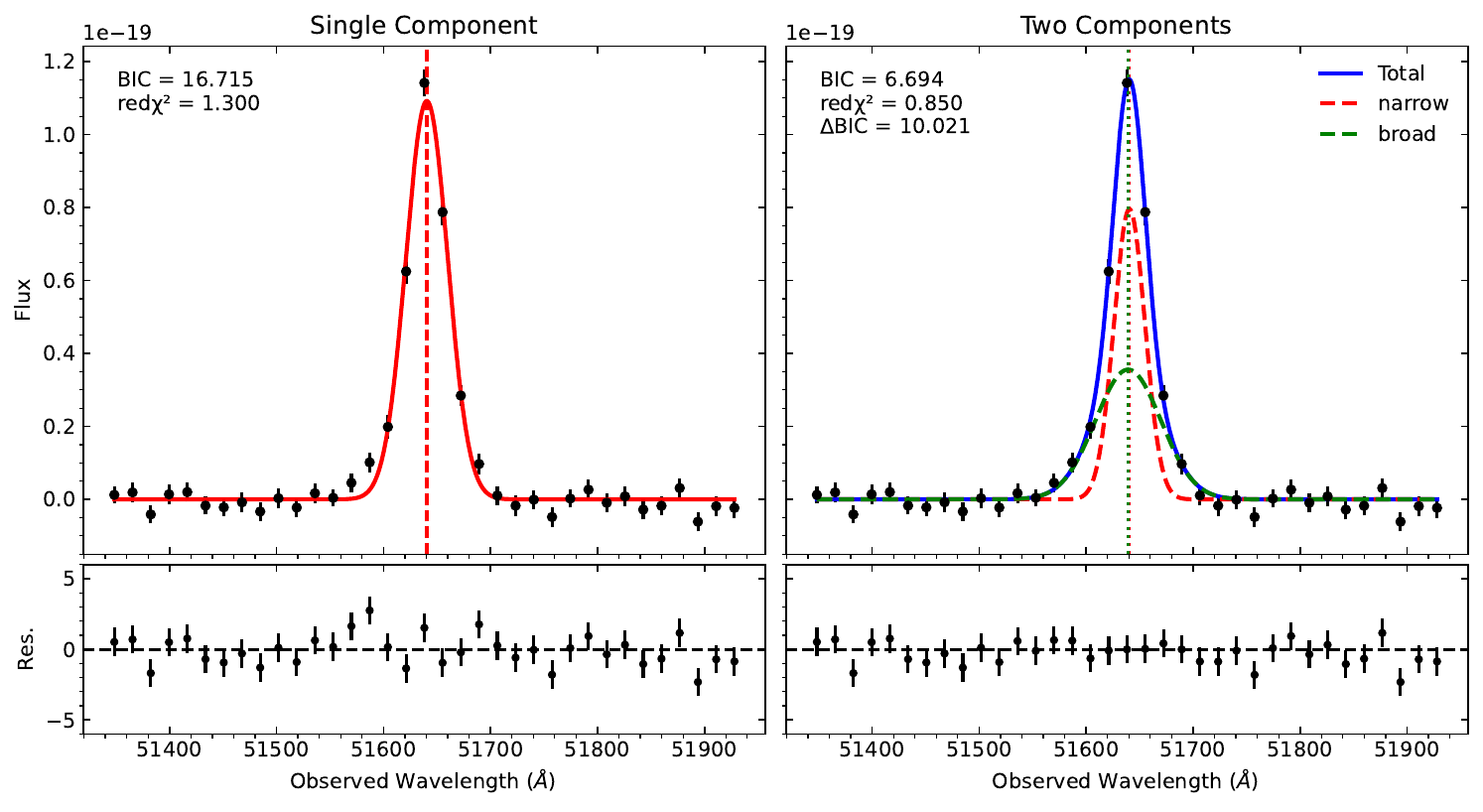}
\caption{[\ion{O}{III}]\,$\lambda5007$ decomposition in the NIRSpec G395M spectrum. The left panel shows the single-Gaussian fit and the right panel the two-component fit. Black points show the spectrum; in the two-component model, blue curve is the total profile, red dashed is the narrow component and green dashed is the broad component. The lower panels show residuals. The two-component model is favored by $\Delta\mathrm{BIC}=10.021$.}
\label{fig:OIII_fit}
\end{figure}

To verify the outflow signature in emission, we analyze the [\ion{O}{III}]\,$\lambda5007$ profile in the G395M spectrum. The line shows extended wings, motivating fits with both single- and two-component Gaussian models using \texttt{lmfit}. The two-component model is favored by $\Delta\mathrm{BIC}\equiv\mathrm{BIC}_{\rm single}-\mathrm{BIC}_{\rm two}=10.021$ (Extended Data Fig.~\ref{fig:OIII_fit}). It separates a narrow component with $F_{\rm narrow}=(2.92\pm0.21)\times10^{-18}\,\mathrm{erg\,s^{-1}\,cm^{-2}}$ and $\mathrm{FWHM}_{\rm narrow}=199\pm23\,\mathrm{km\,s^{-1}}$ from a broad component with $F_{\rm broad}=(2.69\pm0.32)\times10^{-18}\,\mathrm{erg\,s^{-1}\,cm^{-2}}$ and $\mathrm{FWHM}_{\rm broad}=420\pm70\,\mathrm{km\,s^{-1}}$.

Given the independent blueshifted absorption-line signatures in the rest-frame UV spectrum, we interpret the broad [\ion{O}{III}] component as emission associated with ionized outflowing gas. Its small centroid offset relative to the narrow component is consistent with a roughly bipolar geometry in which the approaching and receding sides contribute simultaneously to the spatially integrated profile. The narrow component is therefore used to characterize the systemic, photoionized interstellar medium (ISM).

Using \texttt{PyNeb} \citepmet{Luridiana_pyneb_2015} and only the narrow [\ion{O}{III}] component, we obtain $12+\log(\mathrm{O/H})=8.38\pm0.09$, compared with $8.63\pm0.07$ when the integrated [\ion{O}{III}] flux is used. Inclusion of the broad outflow component would therefore overestimate the ISM oxygen abundance by $\simeq0.25\,\mathrm{dex}$. The abundance calculation, including the adopted electron temperature and density and the associated systematic uncertainties, is described in Supplementary Section~``\hyperref[supp:OIII_abundance]{Gas-phase oxygen abundance and ionization parameter}''.

\subsection*{Physical scale of the outflows}
\label{app:outflow_scale}

We place an approximate upper limit on the distance of the LIS gas from the central star-forming region. If the gas is primarily photoionized by stellar radiation, its ionization parameter is
\begin{equation}
U=\frac{f_{\rm esc}^{\rm ion}Q_{\rm H}}{4\pi R^2n_{\rm H}c},
\label{eq:U}
\end{equation}
where $Q_{\rm H}$ is the production rate of hydrogen-ionizing photons and $f_{\rm esc}^{\rm ion}$ is the fraction reaching the outflow. Guided by radiative-transfer calculations \citepmet{Chisholm_2016}, we adopt $\log U>-3$ only as a weak photoionization-based lower limit. We estimate $Q_{\rm H}$ from the recent SFR \citep{Chen_2026} using the standard conversion of  \citetmet{Kennicutt_1998}, and set $f_{\rm esc}^{\rm ion}=1$ to maximize the inferred radius. These choices give $R<5\,\mathrm{kpc}$; lower escape fractions, higher ionization parameters or higher gas densities reduce this scale. If the HIS gas is dominated by shocks or mixing rather than the same radiation field, the adopted lower limit on $U$ does not apply directly.

The DLA column and fine-structure density also provide an ionization-fraction-dependent upper bound on the line-of-sight thickness of the LIS material. The derivation is given in Supplementary Section~``\hyperref[supp:thickness]{Line-of-sight thickness of the low-ionization gas}''; for equal neutral and ionized hydrogen fractions it gives $\Delta R_{\rm out}<110\,\mathrm{pc}$. This is not a model-independent thickness measurement because the DLA includes all neutral gas along the sightline and the neutral fraction of the outflow is not independently constrained.

\subsection*{Outflow-rate calculation}
\label{app:outflow}

We estimate the ionized mass outflow rate of Gz9p3 using the rest-frame UV absorption lines. Because the electron density is directly constrained from the \ion{C}{II}, \ion{C}{II}$^*$, \ion{Si}{II}, and \ion{Si}{II}$^*$ absorption systems, the mass flux can be estimated without assuming an outflow radius or dynamical time. We approximate the absorbing gas as a uniform flow crossing the projected continuum-emitting area covered by the outflow and calculate
\begin{equation}
\dot{M}_{\rm out}
=
\mu_{\rm H}m_{\rm p}n_{\rm H}v_{\rm out}C_{\rm f}A,
\label{eq:mdot}
\end{equation}
where $A$ is the physical area subtended by the MSA slit on Gz9p3. 
Based on the multi-band imaging morphology, we adopt four-fifths of the slit area. Here, $\mu_{\rm H}=1.4$ is the mean atomic mass per hydrogen nucleus, $m_{\rm p}$ is the proton mass, and $n_{\rm H}$ is the total hydrogen number density inferred from $n_{\rm e}$ following Equation~\ref{eq:nh}.
This formulation assumes that the density and velocity inferred from the absorption-line fit are representative of the gas crossing the covered projected area.

For the fiducial calculation, we adopt $x_{\rm HII}=1$, which minimizes the inferred total hydrogen density and therefore gives a conservative lower limit on the mass outflow rate. The outflow velocity, $v_{\rm out}$, and covering fraction, $C_{\rm f}$, are drawn from the posterior distributions of the joint \ion{C}{II} and \ion{Si}{II} absorption-line fit. Because the measured velocity is projected along the line of sight, $v_{\rm out}$ should be regarded as a conservative estimate of the intrinsic three-dimensional outflow velocity.

We calculate the momentum and energy outflow rates as
\begin{align}
\dot{p}_{\rm out}
&=
\dot{M}_{\rm out}v_{\rm out},
\label{eq:pdot}\\[4pt]
\dot{E}_{\rm out}
&=
\frac{1}{2}\dot{M}_{\rm out}v_{\rm out}^{2},
\label{eq:edot}
\end{align}
and define the mass-loading factor as
\begin{equation}
\eta
\equiv
\frac{\dot{M}_{\rm out}}{\mathrm{SFR}}.
\label{eq:eta}
\end{equation}

The SFR is derived from the H$\beta$ luminosity. Following \citet{Chen_2026}, we adopt negligible dust attenuation, as inferred from the Balmer-line ratios, and do not apply an additional attenuation correction. We convert the H$\beta$ luminosity to H$\alpha$ using the intrinsic Balmer decrement expected for Case B recombination and adopt the H$\alpha$--SFR calibration of \citetmet{Kennicutt_1998}, converted to a Chabrier IMF \citepmet{Chabrier_2003}.

We do not apply a gravitational-lensing correction when calculating $\eta$. The projected area $A$ and the H$\beta$-based SFR have the same dependence on the lensing magnification, $\mu\simeq1.66$ \citep{Boyett_2024}, and the common factor therefore cancels. The inferred mass-loading factor is thus effectively independent of $\mu$, provided that differential magnification between the outflowing and star-forming regions is negligible.

This UV absorption-line method differs from many conventional thin-shell estimates because the directly constrained electron density removes the need to assume an outflow radius, opening angle, or dynamical time \citep{Xu_2022}. However, it measures only gas projected in front of the stellar continuum and therefore preferentially samples the near-side outflow. The broad [\ion{O}{III}] component has little centroid offset relative to the systemic component, suggesting an approximately bipolar geometry. If so, the UV absorption-line estimate underestimates the total galactic outflow rate. Moreover, the calculation includes only the ionized gas traced by the UV absorption lines and does not account for unobserved neutral atomic or molecular phases. The resulting $\dot{M}_{\rm out}$ and $\eta$ should therefore be interpreted as conservative lower limits on the total multiphase outflow.

For comparison, we independently estimate the outflow rate from the broad [\ion{O}{III}] component. Adopting $n_{\rm e}=380\,\mathrm{cm^{-3}}$ gives a mass-loading factor of $\eta^{[\ion{O}{III}]}=0.13^{+0.08}_{-0.06}$. Because the electron density and characteristic radius of the [\ion{O}{III}]-emitting gas are not directly measured, this result depends on the adopted values of $n_{\rm out}$ and $r_{\rm out}$ and is used only as an empirical comparison rather than as the fiducial mass-loading measurement. The complete calculation is described in Supplementary Section~``\hyperref[supp:oiii_outflow]{[\ion{O}{III}]-based outflow calculation}'', and the result is reported in Extended Data Table~\ref{tab:outflow}.
\begin{table*}[htbp]
\centering
\footnotesize
\begin{threeparttable}
\caption{Summary of the physical properties of Gz9p3.}
\label{tab:outflow}
\begin{tabular}{p{0.55\textwidth}c}
\toprule
Physical quantity & Measured value \\
\midrule
\multicolumn{2}{l}{\textit{Emission-line-based estimate}\tnote{a}} \\
H$\beta$-based star-formation rate, $\mathrm{SFR}_{\rm H\beta}/\mathrm{M_\odot\,yr^{-1}}$ & $7.7\pm1.1$ \\
Gas-phase metallicity, $Z/Z_\odot$ & $0.49^{+0.07}_{-0.06}$ \\
Ionization parameter, $\log U$ & $-2.19\pm0.06$ \\
\midrule
\multicolumn{2}{l}{\textit{UV absorption-line-based estimate}} \\
Electron density of the cool LIS outflow, $\log(n_{\rm e}/\mathrm{cm^{-3}})$ & $1.22^{+0.44}_{-0.71}$ \\
Ionized mass outflow rate, $\dot{M}_{\rm out}/\mathrm{M_\odot\,yr^{-1}}$ & $92^{+164}_{-74}$ \\
Ionized momentum outflow rate, $\log(\dot{p}_{\rm out}/\mathrm{dyn})$ & $34.96^{+0.49}_{-0.68}$ \\
Ionized energy outflow rate, $\log(\dot{E}_{\rm out}/\mathrm{erg\,s^{-1}})$ & $41.9^{+0.5}_{-0.7}$ \\
Ionized mass-loading factor, $\eta$ & $12.3^{+21.7}_{-9.9}$ \\
\midrule
\multicolumn{2}{l}{\textit{[\ion{O}{III}]-based empirical estimate}\tnote{b}} \\
{}[\ion{O}{III}]-based $\eta$ with $n_{\rm e}=380\,\mathrm{cm^{-3}}$ & $0.13^{+0.08}_{-0.06}$ \\
\bottomrule
\end{tabular}
\begin{tablenotes}
\footnotesize
\item Uncertainties are propagated with 20,000 Monte Carlo realizations and reported as 68\% HPD intervals estimated with KDE.
\item[a] Emission-line measurements are from \citet{Chen_2026}; only the narrow [\ion{O}{III}] component is used.
\item[b] The empirical estimate adopts $R=1000$ and $Z=0.49\,Z_\odot$; the metallicity is the ISM estimate for Gz9p3.
\end{tablenotes}
\end{threeparttable}
\end{table*}

\subsection*{Escape-velocity estimate}
\label{app:escape_velocity}

The halo mass is not directly measured. We therefore adopt an illustrative range $\log(M_{\rm h}/M_\odot)=10.8$--$11.5$, with a fiducial value of $11.0$, motivated by extrapolated stellar-to-halo mass relations and abundance matching \citepmet{Behroozi_2019} together with high-redshift LBG clustering and HOD constraints \citepmet{Harikane_2018}, broadly consistent with the independent abundance-matching estimate of \citetmet{Aditya_2026}. 

We calculate the virial radius using the Bryan \& Norman (1998) spherical-overdensity prescription \citepmet{Bryan_1998} and estimate
\begin{equation}
v_{\rm esc}(r)=v_{\rm vir}\left[2\left(1+\ln\frac{r_{\rm vir}}{r}\right)\right]^{1/2}
\label{eq:vesc}
\end{equation}
for a truncated isothermal sphere, following common order-of-magnitude outflow comparisons \mainmetcite{Cooper_2025}{Martin_2005,Chisholm_2015}. Across the adopted halo masses, $v_{\rm esc}\simeq350$--$650\,\mathrm{km\,s^{-1}}$ at $2\,\mathrm{kpc}$ and $290$--$560\,\mathrm{km\,s^{-1}}$ at $5\,\mathrm{kpc}$; the fiducial values are $420$ and $350\,\mathrm{km\,s^{-1}}$, respectively. These values are approximate because they depend on halo mass, profile and radius, and they neglect the baryonic potential, which would raise the inner escape speed.

\subsection*{FFB-model prediction and star-formation efficiency}
\label{app:ffb_envelope}

We extend the feedback-free starburst (FFB) wind calculation of \citet{Li_FFB_2024} by incorporating the observed star-formation history (SFH) of Gz9p3. 
We denote by $\varepsilon$ the characteristic star-formation efficiency of the recent burst responsible for driving the observed outflow. 
Following \citet{Li_FFB_2024}, the intrinsic loading follows from mass conservation: a fraction $1-\varepsilon$ of the initial gas remains after the burst, while massive stars promptly return a fraction $f_{\rm ret}=0.2$ of the newly formed stellar mass. Thus,
\begin{equation}
\eta_{\rm int}
=
\frac{1-\varepsilon+f_{\rm ret}\varepsilon}{\varepsilon}
=
\varepsilon^{-1}-0.8 .
\end{equation}

The outflow is observed after it has propagated away from the star-forming region, whereas its apparent loading is normalized by the lower SFR at the epoch of observation. We describe this temporal mismatch by $f_{\rm burst}$, the ratio of the observed SFR to that of the outflow-driving burst. The reconstructed SFH indicates a burst at a lookback time of approximately 10--20 Myr with an SFR about one order of magnitude above that traced by H$\beta$ over the latest ${\sim}10$ Myr \citep{Chen_2026}; we therefore adopt $f_{\rm burst}=0.1$.

We phenomenologically separate the total wind into hot and cool components, $\eta_{\rm int}=\eta_{\rm h}+\eta_{\rm c}$, and adopt $\eta_{\rm h}=0.3$ \citepmet{Strickland_2009}. The apparent cool-phase loading is then
\begin{equation}
\eta_{\rm app}=\frac{\eta_{\rm c}}{f_{\rm burst}}=
\frac{\varepsilon^{-1}-0.8-\eta_{\rm h}}{f_{\rm burst}}.
\label{eq:ffb_eta_apparent}
\end{equation}
For $f_{\rm burst}<1$, a delayed wind can have an apparent loading substantially above the intrinsic instantaneous value. The hot--cool division is not specified by the original FFB calculation and is introduced here to connect the total theoretical wind to the observed multiphase outflow.
We emphasize that the inferred $\eta_{\rm app}$ is inherently dependent on the evolutionary phase at which the system is observed: short-timescale fluctuations in the SFH, particularly observations made after the peak of a burst, can temporarily enhance the apparent loading relative to its intrinsic or time-averaged value.

To display the model in the $(M_\star,\eta)$ plane, we use $M_\star\simeq f_{\rm b}\varepsilon M_{\rm h}$ with $f_{\rm b}=0.16$, sample $\varepsilon=0.1$--$0.5$, and retain solutions with $10.5\leq\log(M_{\rm h}/M_\odot)\leq12$. The halo cut defines the plotted domain rather than a sharp threshold for an individual FFB. For Gz9p3, Equation~\ref{eq:ffb_eta_apparent} is inverted directly: the observed loading constrains $\varepsilon$ for the adopted $f_{\rm burst}$ and $\eta_{\rm h}$, while $M_\star$ is used only to associate the solution with a halo mass. The three plotted markers successively account for the observed one-sided ionized flow, bipolar geometry, and an additional neutral contribution. Because neutral atomic and molecular material can carry a substantial fraction of the outflowing mass \citep{Pandya_2021,Fluetsch_2019,Fluetsch_2021}, the bipolar+neutral marker is the preferred comparison with the predicted cool-phase budget. The detailed construction and momentum-transfer test are given in Supplementary Section~``\hyperref[supp:ffb]{FFB mapping and momentum transfer}''.

% ============================================================================
% Extended Data: these three labels are explicitly cited as Extended Data in
% the existing main text, so they remain here rather than moving to the SI.
% ============================================================================

\begin{figure}[htbp]
\centering
\includegraphics[width=\columnwidth]{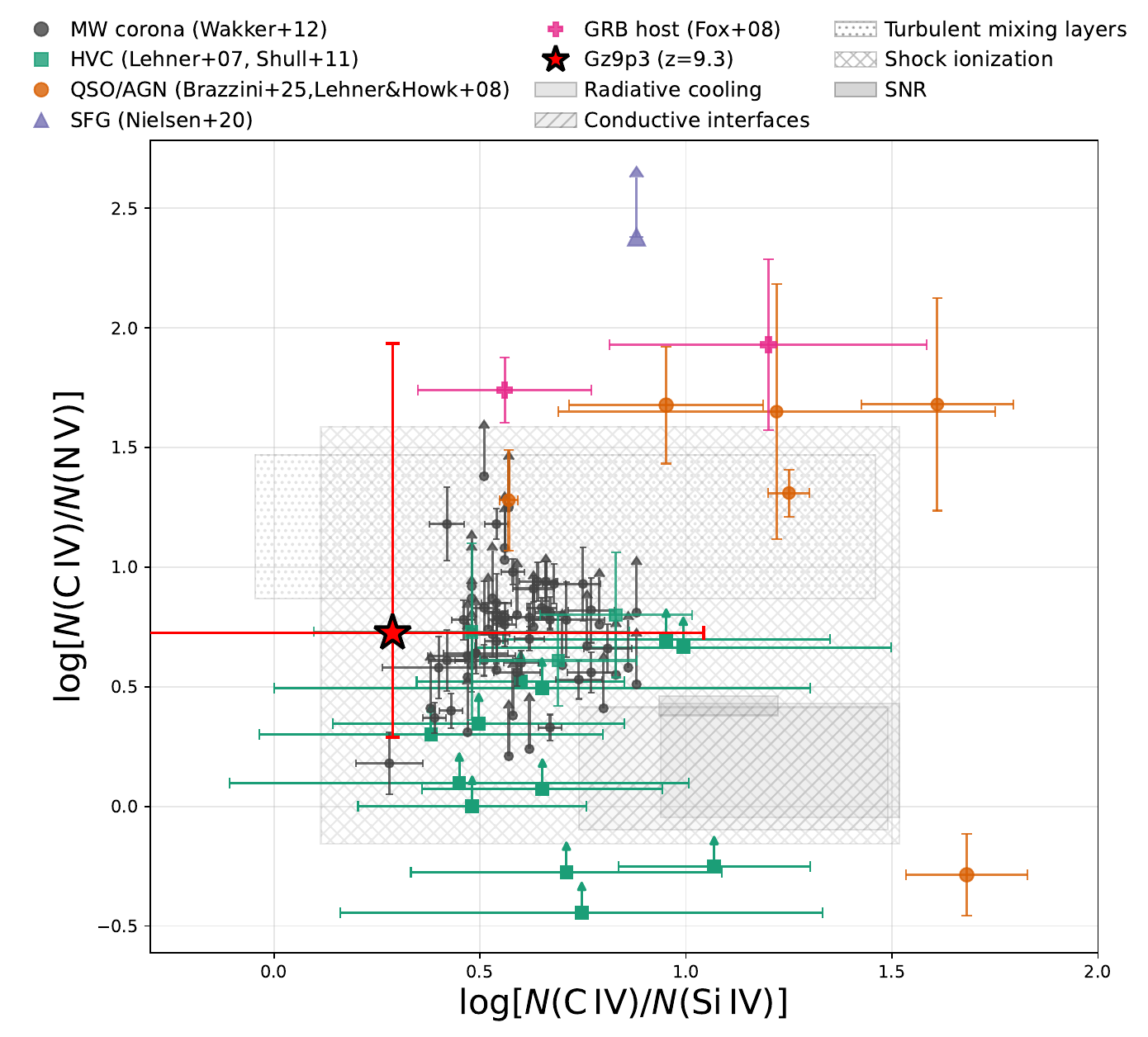}
\caption{\textbf{High-ionization column-density ratios of Gz9p3 compared with local and high-redshift absorbers.} We compare the ionic column-density ratios of Gz9p3 with a compilation of highly ionized absorbers, including the Milky Way corona \protect\citepmet{Wakker_2012}, HVC and intergalactic absorbers \protect\citepmet{Lehner_2007,Shull_2011}, QSO/AGN absorbers \mainmetcite{Brazzini_2025}{Lehner_2008}, star-forming galaxies \protect\citepmet{Nielsen_2020}, and GRB-host absorbers \protect\citepmet{Fox_2008}. The shaded regions show the expected ranges from radiative cooling, conductive interfaces, turbulent mixing layers, shock ionization, and supernova remnants \protect\citepmet{Heckman_2002,Gnat_2007,Borkowski_1990,Slavin_1993,Dopita_1996,Slavin_1992}. Gz9p3 lies within the broad locus occupied by collisionally ionized gas, suggesting that the high-ionization absorption may arise from warm/hot interfaces or shock-heated material rather than a purely photoionized phase.}
\label{fig:HIS_1}
\end{figure}

\begin{figure}[htbp]
\centering
\includegraphics[width=\columnwidth]{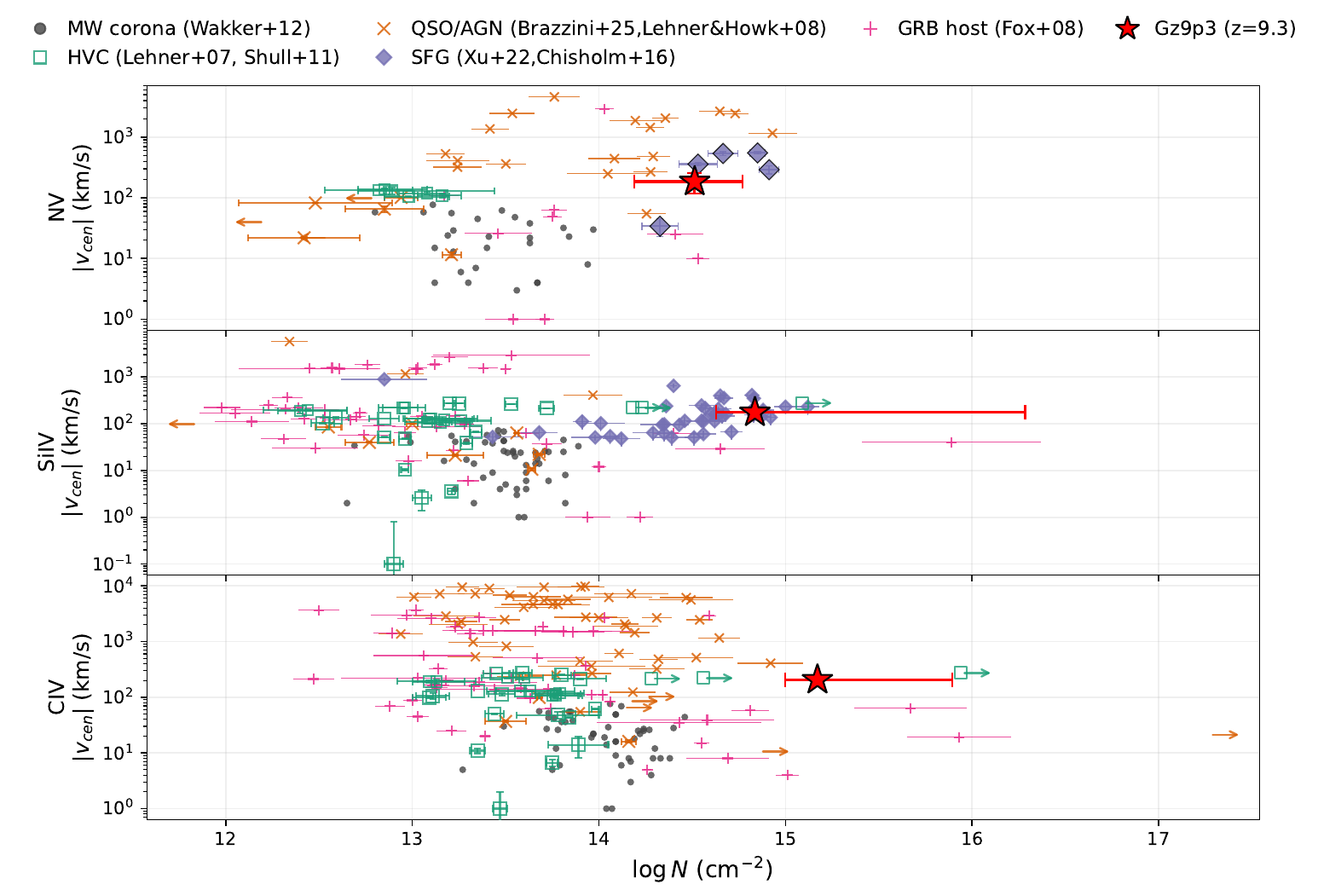}
\caption{\textbf{Column densities and velocity offsets of high-ionization absorbers.} We compare the N\,V, Si\,IV, and C\,IV column densities and centroid velocities of Gz9p3 with measurements from the Milky Way corona \protect\citepmet{Wakker_2012}, high-velocity clouds and intergalactic absorbers \protect\citepmet{Lehner_2007,Shull_2011}, QSO/AGN absorbers \protect\mainmetcite{Brazzini_2025}{Lehner_2008}, UV absorption-line outflows in star-forming galaxies \protect\mainmetcite{Xu_2022}{Chisholm_2016} and GRB-host absorbers \protect\citepmet{Fox_2008}. The column densities shown for Gz9p3 are conservative lower-limit estimates.}
\label{fig:HIS_2}
\end{figure}

\clearpage
% \bibliographystylemet{sn-mathphys-num}
% \bibliographystylemet{Gz9p3}
\bibliographymet{Gz9p3}

\newpage 

\small{
\noindent {\bf Acknowledgments.}
This paper is dedicated to the memory of Prof. Avishai Dekel, whose foundational contributions to galaxy evolution — including cold-mode accretion, violent disk instability, and the feedback-free starburst framework that motivates this work — have profoundly shaped the field.
This work is supported by the National Key R\&D Program of China No.2025YFF0510603, the National Natural Science Foundation of China (grant 12373009), the CAS Project for Young Scientists in Basic Research Grant No. YSBR-062, the China Manned Space Program with grant no. CMS-CSST-2025-A06, and the Fundamental Research Funds for the Central Universities. XW acknowledges the support by the Xiaomi Young Talents Program, and the work carried out, in part, at the Swinburne University of Technology, sponsored by the ACAMAR visiting fellowship. 
KG and TN acknowledge support through the Australian Research Council (ARC) Laureate Fellowship FL180100060. TN also acknowledges support through ARC Discovery Project Grant DP230103161.
% TN thanks support through ARC Discovery Project Grant DP230103161.
We thank the UNCOVER Team (PID-2561, PIs: Labbé and Bezanson) and the SPURS Team (PID-9214, PIs: Mason and Stark) for developing their observing programs with a non-proprietary period.
\\

\noindent {\bf Author contributions.}  
X.W. designed the project. She.W. and X.W. wrote the manuscript.
Ha.Z. reduced the JWST NIRSpec data.
She.W. conducted the detailed fitting of the spectral features and measured the outflow properties.
Z.Q. and Z.L. assisted the interpretation of the main results from the absorption-line analysis.
Y.P. performed the SED fitting with \texttt{Prospector}.
Q.Z. and She.W. conducted the DLA fitting.
Sho.W. performed the X-ray data analysis.
All authors took part in interpreting the results, contributed to the manuscript, and approved the final version.
\\

\noindent {\bf Competing interests.}
The authors declare no competing interests.
\\

\noindent {\bf Data availability.}
The uncalibrated JWST rate products (\texttt{*\_rate.fits}) used in this work are publicly available from the Mikulski Archive for Space Telescopes (MAST; \url{https://mast.stsci.edu/portal/Mashup/Clients/Mast/Portal.html}).
Raw data for the observations taken with Chandra are publicly available through the Chandra Data Archive: \url{https://cda.harvard.edu/chaser/}.
The authors can provide other data supporting the findings of this study upon reasonable request.
\\

\noindent {\bf Code availability.}
Data reduction was performed using the publicly available MSAEXP pipeline (\url{https://github.com/gbrammer/msaexp}). Spectral fitting, DLA modeling, and outflow-parameter calculations were performed using standard open-source packages, including emcee, lmfit, PyNeb, ChiantiPy, and Prospector. Custom analysis scripts are available from the corresponding author upon reasonable request.
}

% ============================================================================
% Supplementary Information
% ============================================================================

\clearpage
\setcounter{figure}{0}
\setcounter{table}{0}
\renewcommand{\figurename}{Supplementary Fig.}
\renewcommand{\tablename}{Supplementary Table}
\renewcommand{\thefigure}{\arabic{figure}}
\renewcommand{\thetable}{\arabic{table}}
\renewcommand{\theHfigure}{SupplementaryFigure.\arabic{figure}}
\renewcommand{\theHtable}{SupplementaryTable.\arabic{table}}

\section*{Supplementary Information}

\subsection*{Ancillary data and additional JWST observations}
\label{supp:ancillary}

We adopt the publicly available reduced JWST/NIRCam mosaics and photometric catalogs released by the UNCOVER team \citepsupp{uncover_mosaic}, which provide multi-band coverage across the available NIRCam filters and are used to characterize the morphology and photometric properties of Gz9p3.

\begin{figure}[htbp]
\centering
\includegraphics[width=0.8\textwidth]{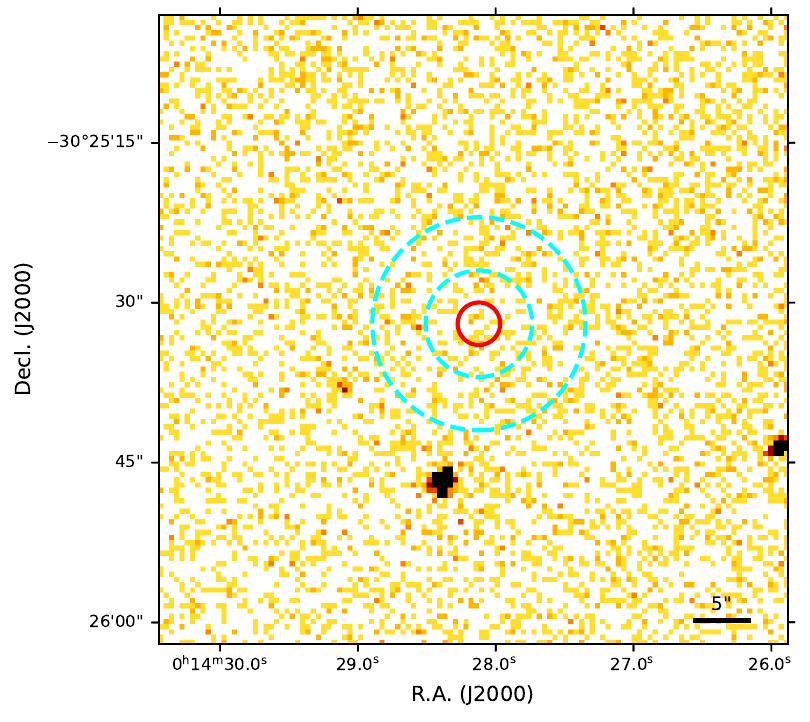}
\caption{Merged 0.5--7 keV Chandra image centered on Gz9p3. The red circle marks the source-extraction aperture, and the cyan annulus indicates the background region.}
\label{fig:xray}
\end{figure}

We search for X-ray emission using 102 merged Chandra observations of the Abell~2744 field. The data reduction follows \citetsupp{Zou_2026}, yielding a total exposure time of $\sim2.2\,\mathrm{Ms}$. Gz9p3 lies at an off-axis angle of $\sim2.2'$, where the 90\% enclosed-energy PSF radius is $\sim1.4''$ \citepsupp{Evans_2024}. We extract source counts with the CIAO tool \texttt{srcflux} using a $2''$-radius aperture centered on the optical position and estimate the background from an annulus with inner and outer radii of $5''$ and $10''$ (Supplementary Fig.~\ref{fig:xray}). No band shows a signal above $3\sigma$. Assuming a power law modified by Galactic absorption with $\Gamma=2$ and $N_{\rm H}=1.34\times10^{20}\,\mathrm{cm^{-2}}$, we obtain a 90\% upper limit $F_X<1.1\times10^{-16}\,\mathrm{erg\,cm^{-2}\,s^{-1}}$ in the 0.5--7 keV band. After correcting for $\mu\simeq1.66$ \citep{Boyett_2024}, the intrinsic limit is $L_X(2\text{--}10\,\mathrm{keV})<4.5\times10^{43}\,\mathrm{erg\,s^{-1}}$ at $z=9.311$. The luminosity limit is sensitive to the photon index; $\Gamma=2.0$ gives a value more than twice that obtained for $\Gamma=1.4$ and is therefore the more conservative assumption.

Gz9p3 has also been observed with NIRSpec IFU PRISM/CLEAR for $2888.6\,\mathrm{s}$ and MIRI MRS in the full-channel/full-band configuration for $19589\,\mathrm{s}$ (GO-4530 \citepsupp{Bik_2026}), and with NIRSpec MSA high-resolution F100LP/\allowbreak G140H, F170LP/\allowbreak G235H, and F290LP/\allowbreak G395H spectroscopy from GLASS-JWST ERS-1324 \citepsupp{Treu_2022}. Because of their different depths, wavelength coverage and observing configurations, these data are not used in the main analysis.

\subsection*{Additional details of the DLA fit}
\label{supp:DLA}

We model the Ly$\alpha$ absorption profiles in the NIRSpec MSA prism and G140M/F100LP spectra using a joint DLA framework following \citet{Chen_2026}. In this high-redshift context, the model includes both local \ion{H}{I} absorption associated with Gz9p3 and attenuation from the neutral IGM along the line of sight. The prism slit crosses the extended stellar-tail structure of Gz9p3, from which we extract spectra at three spatial positions along the slit. These are combined with the G140M/F100LP spectrum of the galaxy center, yielding four spatially resolved sightlines over a projected scale of $2.5\,\mathrm{kpc}$.

For each sightline, the intrinsic rest-frame UV continuum is described by a local power law, $F_\lambda \propto \lambda^\beta$. We also test replacing this power-law continuum with the continuum predicted by the best-fit \texttt{Prospector} SED model \citepsupp{Johnson_prospector_2021}. The inferred $N_{\rm HI}$ values are consistent within the uncertainties, indicating that our results are not driven by the adopted continuum prescription. Because the SED-based continuum constraints vary substantially among the different sightlines, we adopt the same power-law continuum prescription for all extractions. In the forward modeling, each model spectrum is convolved with the instrumental line-spread-function and evaluated on the wavelength grid of the corresponding prism or G140M/F100LP observation, thereby accounting for the different spectral resolutions and sampling of the two datasets.

Since all four sightlines probe the same galaxy, we assume a common IGM neutral fraction, $\bar{x}_{\rm HI}$, with a uniform prior over $[0,1]$, and a common distance from the galaxy to the onset of the neutral IGM along the line of sight, $R_{\rm ion}$. 
At the observed Ly$\alpha$ wavelength, the effective angular resolution is finer than the PRISM spatial-pixel scale, such that the three extractions can be treated as independent spatial sightlines.  Nevertheless, because their projected physical separations within Gz9p3 are small, we assume a common DLA covering fraction, $C_{f,\mathrm{prism}}$, with a uniform prior over $[0.2,1]$ and a common DLA velocity offset, $\Delta v_{\rm DLA,prism}$, with a uniform prior over $[-500,0]\,\mathrm{km\,s^{-1}}$. 
For the galaxy-center sightline observed with G140M/F100LP, we adopt informative priors on $C_{f,\mathrm{G140M}}$ and $\Delta v_{\rm DLA,G140M}$ based on the LIS absorption-line measurements. Together with the four independent $N_{\rm HI}$ values, the model contains ten free parameters. We sample the posterior distributions using MCMC and derive the radial variation of $N_{\rm HI}$ by taking the galaxy center as the origin. We note that the DLA velocity offsets are weakly constrained by the current data and are therefore treated as nuisance parameters. Our interpretation is based on the marginalized $N_{\rm HI}$ distributions rather than on the fitted velocity offsets.

Previous studies have modeled the Ly$\alpha$ break of Gz9p3 using different assumptions for the intrinsic UV continuum, local \ion{H}{I} absorption, and IGM attenuation \mainsuppcite{Chen_2026}{Pollock_2026}. These choices lead to different inferred values of $N_{\rm HI}$, reflecting the degeneracy among continuum placement, the local DLA component, the IGM neutral fraction, and the distance to the neutral IGM. In general, assigning a larger fraction of the Ly$\alpha$ break to local \ion{H}{I} absorption leads to a higher inferred $N_{\rm HI}$, whereas assigning more of the absorption to the IGM damping wing lowers the required local column density. We therefore regard the absolute normalization of $N_{\rm HI}$ as model-dependent.

Our analysis differs from previous work in that we jointly model four spatially resolved Ly$\alpha$ absorption sightlines across the galaxy center and the extended stellar-tail direction. In the joint fit, the four sightlines share the global IGM parameters, while each sightline is assigned an independent local $N_{\rm HI}$. This approach helps separate sightline-to-sightline variations in the local \ion{H}{I} column from the global IGM attenuation, because the IGM parameters are shared among all sightlines.

We emphasize that our main conclusion does not rely on the absolute $N_{\rm HI}$ scale alone. Although different DLA prescriptions may shift the inferred column densities, our spatially resolved fit indicates that the outer prism sightlines along the stellar tail have higher $N_{\rm HI}$ than the central G140M/F100LP sightline. The key result is therefore the spatial gradient in $N_{\rm HI}$ within this joint modeling framework, rather than the exact absolute value of $N_{\rm HI}$ for any individual sightline. This spatial trend supports the interpretation that the extended stellar tail is associated with an enhanced neutral-gas reservoir.

\begin{figure}[htbp]
\centering
\includegraphics[width=0.8\textwidth]{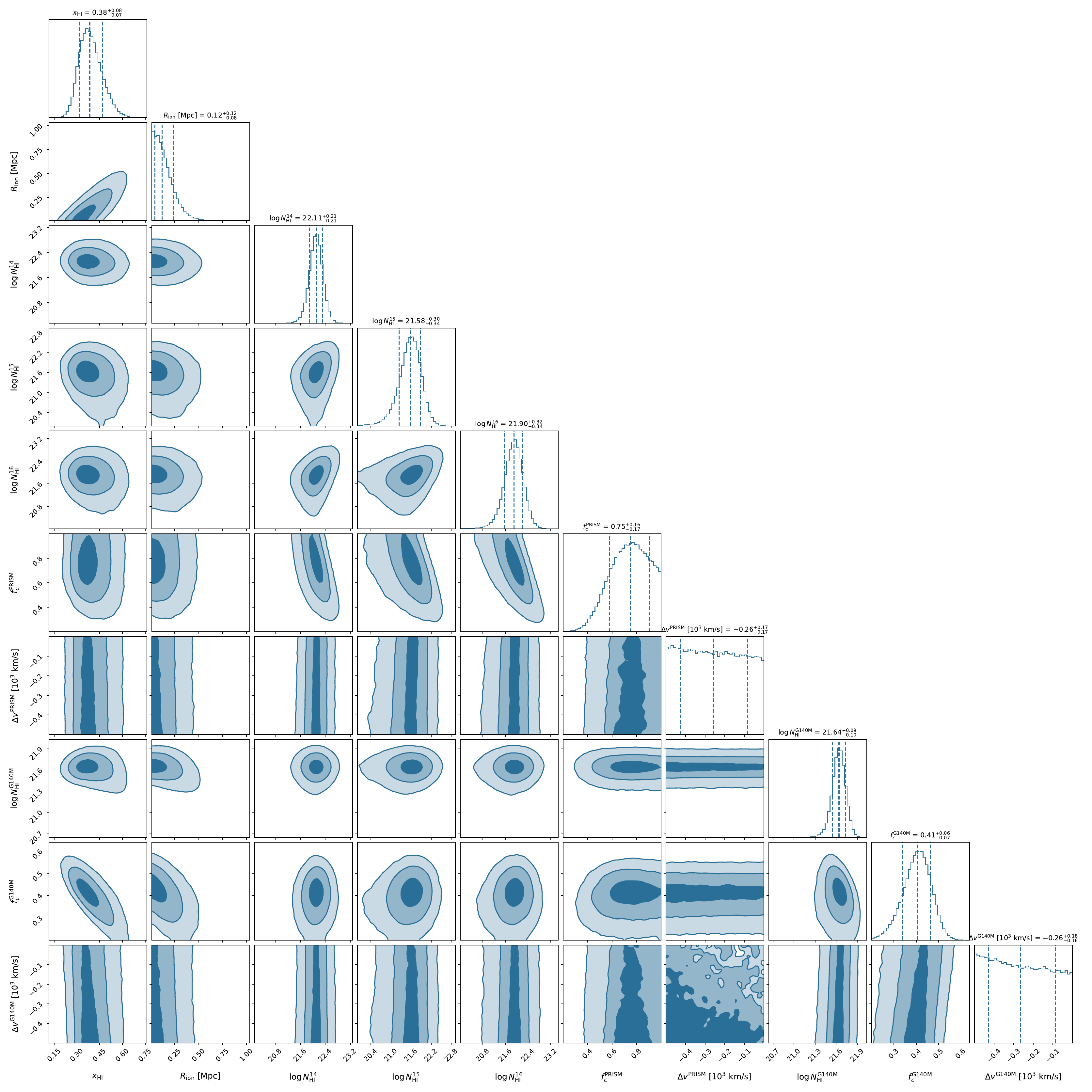}
\caption{Posterior distributions from the joint DLA fit to the four Ly$\alpha$ absorption sightlines in Gz9p3. The model includes $\bar{x}_{\rm HI}$ and $R_{\rm ion}$, while allowing each sightline to have an independent local $N_{\rm HI}$.}
\label{fig:DLA_corner_1}
\end{figure}

\subsection*{Equivalent-width analysis}
\label{supp:EW}

We utilize the clean spectral regions defined in \citetsupp{Rix_2004} to continuum-normalize the rest-frame spectra. EWs of absorption features are then measured on the normalized spectrum within a velocity window of $\pm500\,\mathrm{km\,s^{-1}}$ centered on each line. To place constraints on the electron density of the outflow, we additionally measure $3\sigma$ upper limits on the EWs of three \ion{Si}{II}$^{*}$ fine-structure lines; given potential line blending, a narrower window of $\pm250\,\mathrm{km\,s^{-1}}$ is adopted for these measurements (see Supplementary Table~\ref{tab:abs_lines}).

Based on the measured EWs and oscillator strengths ($f$) of the \ion{Si}{II} and \ion{Si}{IV} absorption lines, we assess their saturation using a curve-of-growth argument. The \ion{Si}{II} transitions span more than an order of magnitude in $f$, whereas their measured EWs differ only by a factor of order unity. Such a weak dependence of EW on $f$ is inconsistent with the optically thin regime of the curve of growth and indicates substantial saturation. A similar comparison of the \ion{Si}{IV} doublet indicates that \ion{Si}{IV}\,$\lambda1393$ is also saturated.

For a saturated absorption line, the residual intensity approaches $I(v)\simeq1-C_{\rm f}$, such that the absorption depth provides a direct constraint on $C_{\rm f}$. Because the intrinsic absorption profiles are unresolved at the spectral resolution of the G140M spectrum ($R\sim1000$), instrumental broadening can make the observed troughs shallower than their intrinsic depths. We therefore use the maximum observed absorption depth of the saturated transitions as a conservative lower limit on $C_{\rm f}$, obtaining $C_{\rm f,LIS}>0.6$ and $C_{\rm f,HIS}>0.8$ (see Supplementary Table~\ref{tab:abs_lines}). A more quantitative constraint on $C_{\rm f}$ is obtained from the joint fine-structure analysis of the \ion{C}{II}, \ion{C}{II}$^*$, \ion{Si}{II}, and \ion{Si}{II}$^*$ absorption systems, in which $C_{\rm f}$ is treated as a free parameter (Methods subsection~``\hyperref[app:CII_ne]{Fine-structure absorption and electron density}'').

We also note that the \ion{Al}{II}\,$\lambda1670$ absorption feature may be blended with nearby \ion{O}{III}] emission. Measurements involving \ion{Al}{II} should therefore be regarded as tentative and are used only as a reference in the following analysis. In addition, \ion{Fe}{II}\,$\lambda1608$ is not significantly detected, with a $3\sigma$ rest-frame upper limit of $\mathrm{EW}<0.6$\text{\AA}. Under the optically thin approximation, this corresponds to a column-density upper limit of $\log[N(\ion{Fe}{II})/\mathrm{cm}^{-2}]<14.65$.

\begin{table*}[htbp]
\centering
\begin{threeparttable}
\caption{Absorption-line measurements.}
\label{tab:abs_lines}
\begin{tabular}{lcccc}
\toprule
Line & $\lambda_{\rm rest}$ ($\text{\AA}$) & $f$ & EW ($\text{\AA}$) & $C_{\rm f,low\,limit}$ \\
\midrule
N\,V          & 1238.821 & 0.157 & $0.7 \pm 0.2$ & --- \\
Si\,II        & 1260.422 & 1.180 & $1.3 \pm 0.1$ & $0.6 \pm 0.1$ \\
Si\,II$^{*}$  & 1264.738 & 1.090 & $<0.3$ & --- \\
O\,I\tnote{a} & 1302.169 & 0.048 & $0.9 \pm 0.1$ & --- \\
Si\,II\tnote{a} & 1304.370 & 0.093 & $0.7 \pm 0.1$ & --- \\
Si\,II$^{*}$  & 1309.276 & 0.080 & $<0.3$ & --- \\
C\,II\tnote{b} & 1334.532 & 0.128 & $1.9 \pm 0.2$ & --- \\
Si\,IV        & 1393.760 & 0.513 & $2.3 \pm 0.2$ & $0.8 \pm 0.1$ \\
Si\,IV        & 1402.773 & 0.254 & $1.5 \pm 0.2$ & --- \\
Si\,II        & 1526.707 & 0.133 & $0.9 \pm 0.2$ & --- \\
Si\,II$^{*}$  & 1533.432 & 0.133 & $<0.5$ & --- \\
C\,IV\tnote{a} & 1548.204 & 0.190 & $2.0 \pm 0.2$ & --- \\
C\,IV\tnote{a} & 1550.781 & 0.095 & $1.4 \pm 0.2$ & --- \\
Fe\,II        & 1608.451 & 0.059 & $<0.6$ & --- \\
Al\,II        & 1670.789 & 1.74 & $0.8 \pm 0.3$ & --- \\
\bottomrule
\end{tabular}
\begin{tablenotes}
\footnotesize
\item EWs are in the rest frame. Covering-fraction lower limits are derived from the residual intensity at line center for saturated transitions. Values preceded by ``$<$'' are $3\sigma$ upper limits. Oscillator strengths are from \citetsupp{Morton_2003} and NIST.
\item[a] The O\,I $\lambda1302$ and Si\,II $\lambda1304$ lines, and the C\,IV doublet, are not fully resolved. Their EWs are measured by forcing the decomposition at the corresponding line centers.
\item[b] C\,II $\lambda1334$ may include C\,II$^*$; we report the combined C\,II+C\,II$^*$ EW.
\end{tablenotes}
\end{threeparttable}
\end{table*}

\subsection*{Additional details of the Voigt-profile fitting}
\label{supp:voigt}

To constrain the ionic column densities and kinematics of the absorbing gas, we perform MCMC-based Voigt-profile fitting of the continuum-normalized spectra. The model parameters include the ionic column density $N$, Doppler parameter $b$, covering fraction $C_{\rm f}$, and velocity offset $\Delta v$.

We first characterize the centroid velocity of each ionic species using a single-velocity-component fit. The resulting $\Delta v$, measured relative to the systemic redshift, is reported in Extended Data Table~\ref{tab:ion}. Given the partial blending between \ion{O}{I}\,$\lambda1302$ and \ion{Si}{II}\,$\lambda1304$, these two transitions are fitted simultaneously. We emphasize that the single-component fits are used only to characterize the bulk velocity of each ionic species and are not intended to recover the full intrinsic velocity structure of the absorbing gas.

The degree to which the remaining absorber properties can be constrained differs among ionic species. For \ion{Si}{II}, multiple transitions spanning more than an order of magnitude in $f$ provide direct evidence for substantial saturation and reduce the degeneracy between $C_{\rm f}$ and optical depth. The residual intensity of the saturated \ion{Si}{II}\,$\lambda1260$ transition therefore provides an empirical lower limit on $C_{\rm f,LIS}$, as described in Supplementary Inforamtion subsection~``\hyperref[supp:EW]{Equivalent-width analysis}''. Similarly, the \ion{Si}{IV} doublet provides evidence for saturation and an empirical lower limit on $C_{\rm f,HIS}$. Nevertheless, at the spectral resolution of the G140M spectrum ($R\sim1000$), instrumental convolution and unresolved velocity substructure prevent a unique reconstruction of the intrinsic $C_{\rm f}$, $b$, and $N$.

For ions represented by only one effectively independent transition or by transitions affected by blending or contamination, the degeneracy among $N$, $b$, and $C_{\rm f}$ is substantially more severe. In particular, \ion{O}{I} is constrained primarily by the \ion{O}{I}\,$\lambda1302$ transition blended with \ion{Si}{II}\,$\lambda1304$, while \ion{Al}{II}\,$\lambda1670$ may be contaminated by nearby \ion{O}{III}] emission. Robust independent constraints on $C_{\rm f}$ are therefore not available for these ions.

Our primary goal in the column-density analysis is therefore not to obtain unique measurements of $N$ for every ion, but to derive conservative lower bounds while preserving physically meaningful comparisons among ions within the same ionization group. For this purpose, we adopt $C_{\rm f}=1$ when deriving the ionic column densities reported in Extended Data Table~\ref{tab:ion}. For a given observed absorption profile, this assumption generally minimizes the column density required by the model and therefore provides a conservative estimate in the presence of unresolved partial covering. The empirical covering-fraction constraints derived from the saturated \ion{Si}{II} and \ion{Si}{IV} transitions are used independently to characterize the spatial coverage of the absorbing gas, whereas the uniform choice of $C_{\rm f}=1$ is adopted specifically to construct a consistent set of conservative ionic column-density estimates.

For the column-density analysis, the LIS transitions (excluding \ion{C}{II}+\ion{C}{II}$^*$) and the HIS transitions are fitted jointly within their respective ionization groups. The ions within each group share a common velocity structure, including the $\Delta v$ and $b$-parameter of each component, while each ionic species has an independent $N$. This approach reduces artificial differences in the inferred ionic column densities that could otherwise arise from independently fitted unresolved velocity structures.

Because the intrinsic absorption profiles may contain multiple narrow components below the spectral resolution, we additionally perform fits allowing two velocity components. These components are not interpreted as uniquely identified physical clouds. Instead, the multi-component fits are used to explore a broader range of unresolved velocity structures and to derive more conservative constraints on the ionic column densities.

\begin{figure}[htbp]
\centering
\includegraphics[width=0.8\textwidth]{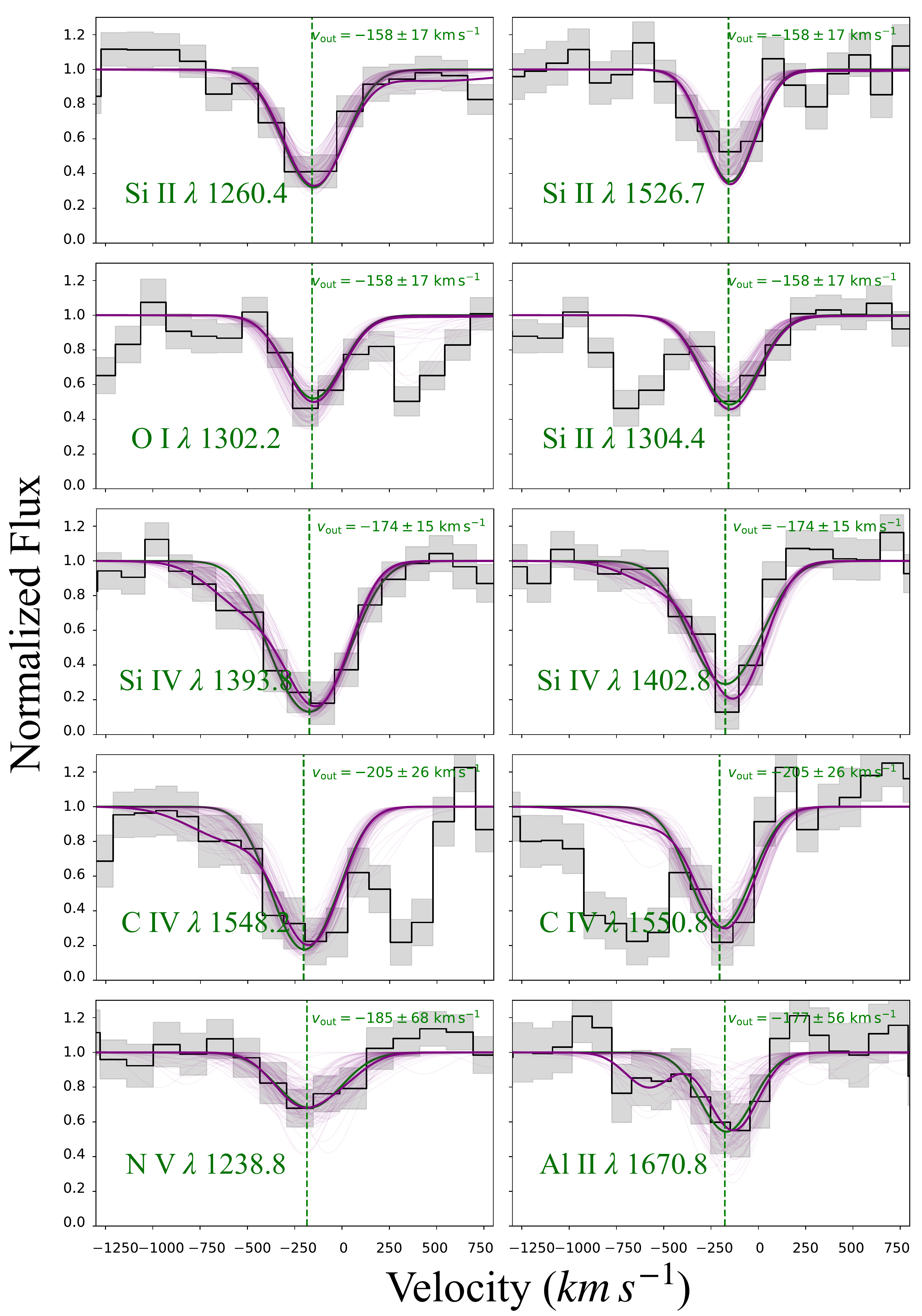}
\caption{MCMC-based Voigt-profile fitting of the rest-frame UV absorption lines. The black histograms and gray regions show the spectra and uncertainties; purple curves show posterior samples and the thick curve the median model. Green dashed lines indicate the best-fit centroid offsets. LIS and HIS features are systematically blueshifted relative to systemic.}
\label{fig:voigt_fitting_result}
\end{figure}

Supplementary Fig.~\ref{fig:voigt_fitting_result} and Extended Data Table~\ref{tab:ion} present the resulting constraints. Posterior distributions are sampled with MCMC using \texttt{emcee}. For each parameter, we report the posterior mode estimated from a Gaussian KDE of the marginalized posterior, together with the 68\% HPD interval. We adopt HPD intervals because several marginalized posteriors are skewed or non-Gaussian.

\subsection*{Outflow $n_{\rm e}$ robustness tests and local comparison}
\label{supp:density}

To assess the robustness of the electron-density constraint, we consider two complementary tests. We first examine whether resonant scattering and fluorescent re-emission could bias the relative depths of the resonance and fine-structure absorption lines. We then apply the same modeling procedure to nearby galaxies with analogous absorption-line properties, providing an empirical check on the method using higher-quality local spectra.

An important source of systematic uncertainty in this method is emission infilling associated with resonant scattering and fluorescent re-emission from \ion{Si}{II} and \ion{C}{II} \citepmet{Gazagnes_2023}. In principle, reprocessed photons emitted at the fine-structure wavelengths can partially fill the corresponding \ion{Si}{II}$^*$ and \ion{C}{II}$^*$ absorption troughs. However, observations of local absorption-selected outflows show that the detected \ion{Si}{II}$^*$ fluorescent emission is generally weak and narrow, and is kinematically distinct from the broad blueshifted absorption associated with the outflow \citepmet{Xu_2023}. A substantial fraction of the reprocessed emission may also arise outside the spectroscopic aperture, since resonantly scattered and fluorescently reprocessed photons can emerge over spatially extended outflow regions \citepmet{Wang_2020_fluo}. In Gz9p3, we detect no significant \ion{Si}{II}$^*$ or \ion{C}{II}$^*$ emission in either the one- or two-dimensional spectra. We therefore model the observed features using absorption components alone and do not explicitly include fluorescent re-emission in our fiducial analysis. Although emission infilling below the current detection limit cannot be completely excluded, the available observations suggest that it is unlikely to dominate the measured absorption profiles.

The combination of strong \ion{C}{II}$^*$ absorption and weak \ion{Si}{II}$^*$ absorption observed in Gz9p3 can also be found in a small number of nearby galaxies. We re-examined the CLASSY sample and identified four galaxies with similar absorption-line properties: J0808+3948 ($z=0.09123$), J1150+1501 ($z=0.002448$), J1225+6109 ($z=0.002341$), and J0337$-$0502 ($z=0.013520$).

After masking foreground Milky Way absorption and contaminating features such as \ion{S}{II}\,$\lambda1259$, we analyzed these four systems using the same forward-modeling procedure adopted for Gz9p3. The inferred electron densities are $\log(n_{\rm e}/{\rm cm^{-3}})=1.26\pm0.07$, $0.81\pm0.11$, $0.69\pm0.12$, and $0.78\pm0.09$, respectively. The derived \ion{Si}{II} column densities are consistent with the measurements reported by \citet{Xu_2022}, while our inferred electron densities are compatible with, and do not contradict, the upper limits reported by \citetmet{Xu_2023}. This agreement demonstrates that, although the method may still be subject to some uncertainty from fluorescent emission infilling, it provides reliable constraints on the electron density of the LIS outflow.

Among the 45 CLASSY galaxies, we identify only four systems exhibiting \ion{Si}{II} and \ion{C}{II} absorption-line characteristics similar to those of Gz9p3, corresponding to an observational fraction of approximately $9\%$. Moreover, J1225+6109 and J0337$-$0502 do not exhibit clear signatures of galactic outflows. Because this fraction depends strongly on observational selection effects, including wavelength coverage, signal-to-noise ratio, line saturation and foreground contamination, it should not be interpreted as the intrinsic occurrence rate of such physical conditions. Nevertheless, it indicates that nearby outflow systems suitable for jointly constraining electron density using the \ion{Si}{II}/\ion{Si}{II}$^*$ and \ion{C}{II}/\ion{C}{II}$^*$ absorption pairs are uncommon. In this context, Gz9p3 represents an exceptional laboratory: not only does it benefit from ultra-deep JWST rest-frame UV spectroscopy, but its outflow electron density also falls within the narrow regime where both \ion{Si}{II} and \ion{C}{II} fine-structure absorption provide simultaneous diagnostic power, allowing the electron density to be constrained despite the low spectral resolution.

\subsection*{Gas-phase oxygen abundance and ionization parameter}
\label{supp:OIII_abundance}

The broad [\ion{O}{III}] component is kinematically distinct from the narrow systemic component and may differ from it in density, ionization state, and chemical abundance. We therefore remove the broad [\ion{O}{III}] contribution when estimating the gas-phase oxygen abundance of the photoionized ISM. We adopt the remaining emission-line measurements from \citet{Chen_2026} and calculate the ionic abundances with \texttt{PyNeb} \citepmet{Luridiana_pyneb_2015}.

No $T_{\rm e}$-sensitive diagnostic line is detected. In particular, the $3\sigma$ upper limit on [\ion{O}{III}]\,$\lambda4363$ implies only $T_{\rm e}\lesssim6\times10^4\,\mathrm{K}$ and is too weak to constrain the abundance. We therefore adopt $T_{\rm e}=10{,}000\,\mathrm{K}$ as a fiducial assumption rather than a direct measurement. A higher electron temperature would yield a lower oxygen abundance.

For the electron density of the systemic ionized ISM, we adopt $n_{\rm e}\simeq225\,\mathrm{cm^{-3}}$, inferred from the [\ion{O}{II}] doublet ratio in the JWST high-resolution NIRSpec MSA spectrum \citep{Boyett_2024}. This density is distinct from the lower fine-structure density measured for the kinematically blueshifted LIS outflow. We also adopt the negligible nebular attenuation inferred from the H$\beta$ and higher-order Balmer-line ratios \citep{Chen_2026}. Because \ion{He}{II} emission is not detected, we do not apply an ionization correction for a possible O$^{3+}$ contribution.

Using only the narrow [\ion{O}{III}] component, we obtain $12+\log(\mathrm{O/H})=8.38\pm0.09$, compared with $8.63\pm0.07$ when the integrated [\ion{O}{III}] flux is adopted. Thus, including the broad outflow component would overestimate the ISM oxygen abundance by $\simeq0.25\,\mathrm{dex}$. Although the absolute abundance remains sensitive to the assumed $T_{\rm e}$ and the contribution of unobserved ionization states, the relative offset introduced by the broad component is largely insensitive to the adopted $T_{\rm e}$.

We further recompute the ionization parameter using the observed $O_{32}$ ratio after excluding the broad [\ion{O}{III}] contribution. Applying the photoionization-model calibration of Kewley \& Dopita (2002) \citepsupp{Kewley_2002} with a model metallicity of $Z=0.49\,Z_\odot$ gives $\log U\simeq-2.19\pm0.06$. This model metallicity is used only for the $O_{32}$--$\log U$ conversion and should not be confused with the oxygen abundance above, whose absolute value depends on the assumed $T_{\rm e}$.

\subsection*{Line-of-sight thickness of the low-ionization gas}
\label{supp:thickness}

The DLA $N_{\rm HI}$ includes all neutral gas along the sightline, not solely outflowing material, and therefore provides only an upper bound. Assuming the LIS gas is approximately uniform, we relate the electron and total hydrogen densities by
\begin{equation}
n_{\rm H}=\frac{n_{\rm e}}{x_{\rm HII}(1+y)},
\label{eq:nh}
\end{equation}
where $x_{\rm HII}=n_{\rm H^+}/n_{\rm H}$ and $y=n_{\rm He}/n_{\rm H}\simeq0.1$. With $n_{\rm HI}=x_{\rm HI}n_{\rm H}$,
\begin{equation}
\Delta R_{\rm out}=\frac{(1+y)N_{\rm HI,out}}{n_{\rm e}}\frac{x_{\rm HII}}{x_{\rm HI}}.
\label{eq:deltaR}
\end{equation}
Because $N_{\rm HI,out}<N_{\rm HI}^{\rm DLA}$, the measured density and column give
\begin{equation}
\Delta R_{\rm out}<110\left(\frac{x_{\rm HII}/x_{\rm HI}}{1}\right)\,\mathrm{pc}.
\end{equation}
The normalization assumes $x_{\rm HI}=x_{\rm HII}=0.5$. The constraint favors a compact structure if the LIS phase retains a substantial neutral fraction but weakens if the gas is overwhelmingly ionized.

This estimate additionally assumes that the DLA column and the electron density inferred from the fine-structure absorption can be associated with the same LIS structure. If they instead arise predominantly from spatially or physically distinct gas phases, the inferred thickness constraint would not apply directly.

\subsection*{[\ion{O}{III}]-based outflow calculation}
\label{supp:oiii_outflow}

For comparison with the fiducial UV absorption-line estimate, we independently estimate the ionized outflow rate from the broad [\ion{O}{III}]\,$\lambda5007$ component. We follow the empirical prescription of \citet{Cooper_2025} and define the outflow velocity as
\begin{equation}
v_{\rm out}^{[\ion{O}{III}]}
=
|v_{\rm broad}-v_{\rm narrow}|
+
2\left(
\sigma_{\rm broad,obs}^{2}
-
\sigma_{\rm LSF}^{2}
\right)^{1/2},
\label{eq:oiii_vout}
\end{equation}
where $v_{\rm broad}-v_{\rm narrow}$ is the centroid-velocity difference between the broad and narrow components, $\sigma_{\rm broad,obs}$ is the observed velocity dispersion of the broad component, and $\sigma_{\rm LSF}$ is the instrumental line-spread-function width.

We infer the ionized gas mass from the luminosity of the broad [\ion{O}{III}] component following \citetsupp{Carniani_2015}:
\begin{equation}
M_{\rm out}^{[\ion{O}{III}]}
=
0.8\times10^{8}
\left(
\frac{L_{[\ion{O}{III}]}}
{10^{44}\,\mathrm{erg\,s^{-1}}}
\right)
\left(
\frac{Z_{\rm out}}
{Z_\odot}
\right)^{-1}
\left(
\frac{n_{\rm out}}
{500\,\mathrm{cm^{-3}}}
\right)^{-1}
M_\odot.
\label{eq:oiii_mout}
\end{equation}
Here, $L_{[\ion{O}{III}]}$ is the luminosity of the broad component, $Z_{\rm out}$ is the metallicity of the emitting gas, and $n_{\rm e}$ is its electron density. We adopt $Z_{\rm out}=0.49\,Z_\odot$ and $n_{\rm out}=380\,\mathrm{cm^{-3}}$, following the value commonly assumed for high-redshift emission-line outflows \citep{Carniani_2024}. The adopted density is not directly measured for Gz9p3 and constitutes one of the main systematic uncertainties in this calculation. In particular, the inferred gas mass and mass outflow rate both scale inversely with $n_{\rm out}$.

The mass outflow rate is calculated as
\begin{equation}
\dot{M}_{\rm out}^{[\ion{O}{III}]}
=
\frac{
M_{\rm out}^{[\ion{O}{III}]}
v_{\rm out}^{[\ion{O}{III}]}
}
{r_{\rm out}},
\label{eq:oiii_mdot}
\end{equation}
where $r_{\rm out}$ is the characteristic radius of the emitting outflow. We adopt $r_{\rm out}=2\,\mathrm{kpc}$, approximately corresponding to the projected extent of the outflowing region. Because the broad [\ion{O}{III}] emission is not sufficiently spatially resolved to measure this radius directly, the inferred mass outflow rate retains an inverse dependence on the adopted $r_{\rm out}$.

Using the same H$\beta$-based SFR adopted for the UV absorption-line calculation, we obtain a mass-loading factor of
\[
\eta^{[\ion{O}{III}]}
=
0.13^{+0.08}_{-0.06}.
\]
Both the [\ion{O}{III}]-based mass outflow rate and the H$\beta$-based SFR scale linearly with intrinsic luminosity, so their ratio is effectively independent of the lensing magnification provided that differential magnification is negligible.

Unlike the UV absorption lines, which preferentially sample outflowing gas projected in front of the stellar continuum, the broad [\ion{O}{III}] emission can receive contributions from both the approaching and receding sides of the outflow. The inferred value therefore characterizes the spatially integrated O$^{++}$-emitting phase. It does not include ionized gas in other ionization states or neutral atomic and molecular material and should not be interpreted as the total multiphase mass-loading factor.

The result also depends on the adopted velocity prescription and geometry. The quantity $|v_{\rm broad}-v_{\rm narrow}|+2\sigma_{\rm broad,int}$ characterizes the high-velocity wings of the spatially integrated line profile but does not uniquely determine the three-dimensional bulk velocity. Similarly, Equation~\ref{eq:oiii_mdot} represents the emitting gas using a single characteristic radius and flow time. Given these assumptions, particularly the unmeasured electron density and radius of the [\ion{O}{III}]-emitting phase, we treat the [\ion{O}{III}]-based mass-loading factor as an empirical comparison rather than as the fiducial outflow measurement. The result is reported in Extended Data Table~\ref{tab:outflow}.

\subsection*{FFB mapping and momentum transfer}
\label{supp:ffb}

Under the short-SFH approximation, the stellar and halo masses are related by
\begin{equation}
M_\star\simeq f_{\rm b}\varepsilon M_{\rm h},
\label{eq:ffb_mstar_mapping}
\end{equation}
where $f_{\rm b}=0.16$ is the cosmic baryon fraction. A fixed $\varepsilon$ therefore gives a horizontal line in the $(M_\star,\eta)$ plane; stellar mass enters through the inferred halo mass and the adopted domain over which the model is displayed.

The green region in the main figure is constructed by sampling $\varepsilon=0.1$--$0.5$ and retaining solutions with
\begin{equation}
10.5\leq\log(M_{\rm h}/M_\odot)\leq12,
\end{equation}
where the upper halo-mass limit of $10^{12}\,M_\odot$ is set by the maximum halo mass explicitly explored in the calculations of \citet{Li_FFB_2024}.
At low stellar masses, high-efficiency solutions can correspond to $M_{\rm h}<10^{10.5}\,M_\odot$ and therefore fall outside the adopted calculation range.

The three Gz9p3 markers illustrate successive systematic corrections to the inferred mass flux rather than independent measurements. They represent the directly observed one-sided ionized outflow, a bipolar ionized outflow, and a bipolar outflow including an additional neutral contribution. The inferred star-formation efficiency remains conditional on the adopted $f_{\rm burst}$, $\eta_{\rm h}$, outflow geometry, and completeness of the observed gas phases. It should therefore be interpreted as a constraint within the adopted FFB framework rather than as a unique statistical determination.

To assess whether the observed cool outflow can be dynamically coupled to the feedback-driven wind, we consider a simple two-phase momentum-transfer picture. Stellar feedback first thermalizes part of its mechanical energy and drives a fast, low-density hot wind. Interactions between this hot wind and the surrounding cooler gas then transfer a fraction of the hot-wind momentum to the cool outflow traced by the UV absorption lines. We denote this momentum-transfer fraction by $\varphi_p$. Because the velocity and mass loading of the cool phase are observationally constrained, this relation can be inverted to estimate the momentum-transfer efficiency required to produce the observed outflow.

Following \citet{Li_FFB_2024}, energy conservation for the thermalized hot wind gives a characteristic velocity
\begin{equation}
v_{\rm h}
=
1490
\sqrt{\frac{\alpha_{\rm th}}{\eta_{\rm h}}}
\,\mathrm{km\,s^{-1}},
\label{eq:ffb_vhot}
\end{equation}
where $\alpha_{\rm th}$ is the fraction of the available stellar-feedback mechanical energy that is thermalized into the hot wind. The scaling $v_{\rm h}\propto(\alpha_{\rm th}/\eta_{\rm h})^{1/2}$ follows directly from equating the kinetic-energy flux of the hot wind to the thermalized feedback energy injection.

We then define $\varphi_p$ as the fraction of the hot-wind momentum flux transferred to the cool phase,
\begin{equation}
\varphi_p
\equiv
\frac{\dot{p}_{\rm c}}{\dot{p}_{\rm h}}
=
\frac{\eta_{\rm c}v_{\rm out}}
{\eta_{\rm h}v_{\rm h}}.
\end{equation}
Using $\eta_{\rm c}=f_{\rm burst}\eta_{\rm app}$ gives
\begin{equation}
\varphi_p
=
\frac{
f_{\rm burst}\eta_{\rm app}v_{\rm out}
}{
1490\,\mathrm{km\,s^{-1}}
\sqrt{\alpha_{\rm th}\eta_{\rm h}}
}.
\label{eq:ffb_phi_momentum}
\end{equation}

Adopting $v_{\rm out}=160\,\mathrm{km\,s^{-1}}$, $f_{\rm burst}=0.1$, $\alpha_{\rm th}=1$, and $\eta_{\rm h}=0.3$, the one-sided ionized estimate with $\eta_{\rm app}=12.3$ requires $\varphi_p=0.24$. Correcting for a bipolar geometry gives $\eta_{\rm app}=24.6$ and $\varphi_p=0.48$, whereas additionally including the neutral contribution gives $\eta_{\rm app}=49.2$ and $\varphi_p=0.96$.

The bipolar+neutral case therefore approaches complete transfer of the available hot-wind momentum. Because the measured outflow velocity is a lower limit to the intrinsic three-dimensional velocity, the inferred momentum-transfer fractions are also lower limits for the adopted $\alpha_{\rm th}$ and $\eta_{\rm h}$. The physical requirement $\varphi_p\leq1$ consequently leaves little additional momentum margin for the bipolar+neutral interpretation.

\clearpage
% \bibliographystylesupp{sn-mathphys-num}
\bibliographysupp{Gz9p3}

\end{document}